\documentclass{aa}
\pdfoutput=1
\usepackage[english]{babel}
\usepackage[utf8]{inputenc}
\usepackage[varg]{txfonts}
\usepackage{amsmath}
\usepackage{amssymb}
\usepackage{graphicx}
\usepackage{natbib,twoopt}
\usepackage{caption}
\usepackage[breaklinks=true]{hyperref} 
\bibpunct{(}{)}{;}{a}{}{,} 
\makeatletter
\newcommandtwoopt{\citeads}[3][][]{\href{http://adsabs.harvard.edu/abs/#3}%
{\def\hyper@linkstart##1##2{}%
\let\hyper@linkend\@empty\citealp[#1][#2]{#3}}}
\newcommandtwoopt{\citepads}[3][][]{\href{http://adsabs.harvard.edu/abs/#3}%
{\def\hyper@linkstart##1##2{}%
\let\hyper@linkend\@empty\citep[#1][#2]{#3}}}
\newcommandtwoopt{\citetads}[3][][]{\href{http://adsabs.harvard.edu/abs/#3}%
{\def\hyper@linkstart##1##2{}%
\let\hyper@linkend\@empty\citet[#1][#2]{#3}}}
\newcommandtwoopt{\citeyearads}[3][][]%
{\href{http://adsabs.harvard.edu/abs/#3}
{\def\hyper@linkstart##1##2{}%
\let\hyper@linkend\@empty\citeyear[#1][#2]{#3}}}
\makeatother

\title{Reconstructing the galaxy density field with photometric redshifts: I. Methodology and validation on stellar mass functions}

\author{
N. Malavasi \inst{1}\thanks{\textit{E-mail contact:} 
\href{mailto:nicola.malavasi@unibo.it}{nicola.malavasi@unibo.it}} 
\and
L. Pozzetti \inst{2}
\and
O. Cucciati \inst{1}$^,\,$ \inst{2}
\and
S. Bardelli \inst{2}
\and 
A. Cimatti \inst{1}}

\institute{
University of Bologna, Department of Physics and Astronomy (DIFA), v.le Berti Pichat 6/2 - 40127 Bologna, Italy 
\and 
INAF--Osservatorio Astronomico di Bologna, via Ranzani 1 - 40127 Bologna, Italy}

\date{Received <date> / Accepted <date>}

\abstract{Measuring environment for large numbers of galaxies in the distant Universe is an open problem in astrophysics, as environment is important in determining many properties of galaxies during their formation and evolution. In order to measure galaxy environments, we need galaxy positions and redshifts. Photometric redshifts are more easily available for large numbers of galaxies, but at the price of larger uncertainties than spectroscopic ones.}
{In this work we study how photometric redshifts affect the measurement of galaxy environment and how the reconstruction of the density field may limit an analysis of the galaxy stellar mass function (GSMF) in different environments.}
{Using mock galaxy catalogues, we measured the environment with a fixed aperture method, using each galaxy's true and photometric redshifts. We varied the parameters defining the fixed aperture volume, exploring different configurations. We also used photometric redshifts with different uncertainties to simulate the case of various surveys. We then computed GSMF of the mock galaxy catalogues as a function of redshift and environment to see how the environmental estimate based on photometric redshifts affects their analysis.}
{We found that only when using high-precision photometric redshifts with $\sigma_{\varDelta z/(1+z)} \lesssim 0.01$, the most extreme environments can be reconstructed in a fairly accurate way, with a fraction $\ge 60\div 80\%$ of galaxies placed in the correct density quartile and a contamination of $\le 10\%$ by opposite quartile interlopers. A length of the volume in the radial direction comparable to the $\pm 1.5\sigma$ error of photometric redshifts and a fixed aperture radius of a size similar to the physical scale of the studied environment grant a better reconstruction than other volume configurations. When using such an estimate of the density field, we found that any differences between the starting GSMF (divided accordingly to the true galaxy environment) will be damped on average of $\sim 0.3$ dex when using photometric redshifts, but will be still resolvable. Although derived with mock galaxy catalogues, these results may be used to interpret real data as we obtained them by comparing results between the true redshift and photometric redshift case, therefore in a way that does not completely depend on how well the mock catalogues reproduce the real galaxy distribution.}
{This work allows several useful considerations on how to interpret results of an analysis of the GSMF in different environments when the density field is measured using photometric redshifts and represent a preparatory study for future wide area photometric redshift surveys such as the Euclid Survey. We plan to apply the result of this work to an environmental analysis of the GSMF in the UltraVISTA Survey in future work.}

\keywords{Methods: data analysis - Galaxies: luminosity function, mass function - Galaxies: statistics - Galaxies: distances and redshifts - Galaxies: clusters: general}

\titlerunning{Methodology and validation on stellar mass functions}

\begin{document}

\maketitle

\section{Introduction}
\label{intro}
It is a current view in modern astrophysics that the formation and evolution of galaxies are influenced by the environment in which they reside. Although the actual specific effects of the environment on galaxy properties are still largely debated, it is not possible to neglect the importance that the density field has in determining many galaxy parameters. Star formation quenching is thought to be deeply related to environment (see \textit{e.g.} \citeads{2010ApJ...721..193P}, \citeads{2013A&A...556A..55I}) through a variety of physical processes \citepads[see \textit{e.g.} figure 10 
of][]{2003ApJ...591...53T}. As a full understanding of the role of the environment is still missing, the study of the density field in detail is vital for creating a consistent picture of galaxy evolution. 

Many observational evidences have been gathered (both at low and high redshift) of environment having effects on all the main observables, from galaxy colors (see \textit{e.g.} \citeads{1980ApJ...236..351D}, \citeads{2004ApJ...615L.101B}, \citeads{2006MNRAS.373..469B}, \citeads{2006A&A...458...39C}, \citeads{2010A&A...524A...2C}, \citeads{2006MNRAS.370..198C}, \citeads{2007MNRAS.376.1445C}, \citeads{2011MNRAS.411..675S}, \citeads{2012ApJ...746..188M}, \citeads{2013ApJS..206....3S}), to galaxy scaling relations \citepads[see \textit{e.g.}][and references therein]{2012ApJ...756..117F}, radio AGN presence (see \textit{e.g.} \citeads{2010A&A...511A...1B}, \citeads{2014MNRAS.445..280H}, and \citeads{2015A&A...576A.101M}) and galaxy masses and star formation activity. These two last properties are better analysed through the use of the galaxy stellar mass function (GSMF). A key feature of environmental processes is that their effect is visible in the GSMF. A differential analysis based on environmental density of the galaxy stellar mass function may thus help unveiling the physical processes that lead galaxies to form and evolve.

For this reason, many studies have focused on the relation between GSMF and environment, analysing this problem both at $z \lesssim 1.5$ (see \textit{e.g.} \citeads{2001ApJ...557..117B}, \citeads{2003MNRAS.346....1K}, \citeads{2006ApJ...651..120B}, \citeads{2010MNRAS.409..337C}, \citeads{2010A&A...524A..76B}, Davidzon et al., in prep.) and higher (for example \citeads{2015MNRAS.447....2M}, \citeads{2015ApJ...805..121D}). Although qualitatively in agreement among them, these works rely on different methods to estimate environments, with various degrees of accuracy. This is not surprising, as a universal definition of environment is still missing and different methods probe different spatial scales through the use of different tracers (for brief reviews of the various adopted methods in the literature see \textit{e.g.} \citeads{2010A&A...524A...2C}, \citeads{2010ApJ...708..505K}, \citeads{2012MNRAS.419.2670M}, \citeads{2015ApJ...805..121D} and references therein).

The first studies of the GSMF in different environments (for example \citeads{2001ApJ...557..117B}, \citeads{2003MNRAS.346....1K}, \citeads{2006ApJ...651..120B}, \citeads{2010MNRAS.409..337C}, \citeads{2010A&A...524A..76B}, Davidzon et al., in prep.) all relied on spectroscopic redshifts to determine the density field and are all limited to $z \le 1.5$. Spectroscopic redshifts indeed grant small errors and high accuracy in the estimate of galaxy positions, but are not available for faint sources and have a limited redshift range (usually not larger than $0 < z < 1.5$, spectroscopic redshifts may be available at higher redshifts but for very small sky areas). Moreover, given the flux limit and the small sky coverage, sampling rate for spectroscopic redshifts is tipycally low.

In recent years, large and new datasets have become available as wide-area and high-redshift sky surveys have been undertaken and many more are planned (\textit{e.g.} COSMOS, \citeads{2007ApJS..172....1S}, \citeads{2009ApJ...690.1236I}, UltraVISTA, \citeads{2012A&A...544A.156M}, \citeads{2013A&A...556A..55I}, UKIDSS, \citeads{2007MNRAS.379.1599L} among those for which photometric redshifts are available, J-PAS, ALHAMBRA, DES, and Euclid among those ongoing or planned). This great availability of data has been made possible through the use of photometric redshifts, which allow to probe the galaxy population on wider areas and at higher redshifts compared to spectroscopic ones. Recent studies have increasingly relied on photometric redshifts to study the GSMF in different environments (see \textit{e.g.} \citeads{2015MNRAS.447....2M}, \citeads{2015ApJ...805..121D}).

Unfortunately, photometric redshifts lack of the precision offered by spectroscopic samples and this limits their potential. Therefore, the issue of using photometric redshifts to estimate galaxy environments needs to be carefully investigated, as a high uncertainty in the redshift measurement may lead to an inaccurate environmental estimate, biasing an interpretation of the GSMF in relation to environment. This has already been studied at low redshifts by \citetads{2005ApJ...634..833C}, \citetads{2015arXiv150501171E} and has been also marginally investigated in \citetads{2012MNRAS.419.2670M} and \citetads{2015MNRAS.446.2582F}. There are also some works where photometric redshifts are used in synergy with spectroscopic redshifts for environmental studies (see \textit{e.g.} \citeads{2010ApJ...708..505K}, \citeads{2014A&A...565A..67C}, and \citeads{2015A&A...576L...6S}).

In this work we explore how the reconstruction of the environment is affected by the use of photometric redshifts, pushing the analysis to $z \sim 3$ and studying in detail the effect on the GSMF. Using simulated data we compare the density field as obtained with each galaxy's true redshifts and photometric ones. We then apply our analysis to GSMF as a function of redshift and galaxy environment.

We describe the galaxy mock catalogue used in this work in section \ref{data}, while we explain the method used to measure galaxy environments in section \ref{method}. We discuss results on the environmental reconstruction for the best-case photometric redshift errors in section \ref{results}, and we study the impact of varying the photometric redshift uncertainty in section \ref{photozerror}. We finally investigate the dependence of the GSMF on the accuracy of the environmental reconstruction for the best-case photometric redshift error in section \ref{mfrec}. We summarize conclusions in section \ref{conclusions}. Throughout this work we will assume the standard cosmology ($\varOmega_m = 0.3$, $\varOmega_{\varLambda} = 0.7$, $H_0 = 70\: km\cdot s^{-1} Mpc^{-1}$).

\section{Mock Data}
\label{data}
In this work we aim to determine the effect of photometric redshifts on the study of galaxy stellar mass functions in different environments. A possible way to achieve this would be to compare the measurement of the density field for a sample of galaxies with both measured spectroscopic redshifts and measured or simulated photometric ones, taking the environmental estimate in the case when spectroscopic redshifts are used as a reference. This has been done for example by \citeads{2015arXiv150501171E} using SDSS galaxies. In our work we chose a different approach, by relying on mock galaxy catalogues. In this way we extended our analysis to redshifts higher than $z > 1.5$ while considering a large sky area to increase sample statistic. Moreover, by comparing the density field and GSMF when using each galaxy's true and photometric redshifts we were able to derive results that do not completely depend on how well the mock galaxy catalogues reproduce the spatial distribution of real galaxies. In this way, the results of this work can be applied also to real data, as we plan to do in future work.

The dataset that we used is composed by the mock galaxy catalogues of \citetads{2013MNRAS.429..556M}\footnote{The mock catalogues are freely available for download at \url{http://astro.dur.ac.uk/~d40qra/lightcones/EUCLID/}}. These were created for the Euclid Survey, they cover an area of $100\:\deg^2$ in the redshift range $z \in [0,3]$, and are limited to a maximum magnitude of $H \le 27$. They were constructed using the \emph{Millennium Run} dark matter simulation \citepads{2005Natur.435..629S} and the \textsc{galform} semi-analytic model of galaxy formation (\citeads{2000MNRAS.319..168C}, \citeads{2006MNRAS.370..645B}, \citeads{2012MNRAS.426.2142L}). A complete description of the mock lightcones can be found in \citetads{2013MNRAS.429..556M}. A summary of the lightcone geometry and physical parameters can be found in Table \ref{cosmology}.

\begin{table}
\caption{Lightcone geometry and cosmology.}
\label{cosmology}
\centering
\begin{tabular}{c c}
\hline\hline
Parameter & Value \\
\hline
$\varOmega_m$				&  0.25      \\
$\varOmega_{\varLambda}$		&  0.75      \\
$\varOmega_b$				&  0.045     \\
$H_0\: (km\cdot s^{-1}\cdot Mpc^{-1})$	&  73        \\
Redshift range				&  0.0-3.0   \\
Sky coverage $(\deg^2)$			&  100       \\
Field centre (R.A.,Dec) $(\deg)$	&  (0.0,0.0)   \\
Maximum H-band magnitude		&  27        \\
\hline
\end{tabular}
\end{table}

For our purposes, we did not use the whole lightcone, but rather we extracted a smaller square area of $8\:\deg^2$ from the whole $100\:\deg^2$. We tuned several lightcone parameters so to match at best the UltraVISTA Survey (see \citeads{2012A&A...544A.156M} for the survey overview and data reduction process, and \citeads{2013A&A...556A..55I} for the photometric redshift calculation). We kept the redshift range unaltered, but we introduced a further cut in K-band magnitude to $K \le 24$ to be consistent with the UltraVISTA data magnitude limit and with the one expected for the Euclid Survey. The final sample is composed of 1\,054\,752 galaxies.

For each galaxy, several parameters were available, but in particular we used two redshifts values in the estimate of the density field. One is the true (cosmological) redshift of each galaxy ($z_{true}$) and the other is the same redshift to which the peculiar motion of each galaxy was added ($z_{obs}$). In order to create a photometric redshift measure for each source, we randomly extracted values from a Gaussian distribution with dispersion $\sigma_{\varDelta z/(1+z)}\times (1+z_{obs})$ and we added them to each galaxy's $z_{obs}$. In the following we will refer to these constructed photometric redshifts as $z_{phot}$.

\section{The measurement of the density field}
\label{method}
Although several ways of estimating the local environment around a given source are available (see \emph{e.g.} \citeads{2010A&A...524A...2C}, \citeads{2010ApJ...708..505K}, \citeads{2012MNRAS.419.2670M}, \citeads{2015ApJ...805..121D}), in this work we applied only a fixed aperture method, as this approach allows to choose a scale for the environment parametrization which is independent of redshift. In the following, we will refer to the \textit{True} environment ($\varrho_{true}$) when the $z_{true}$ of each galaxy is used for the environmental estimate, while we will refer to the \textit{Reconstructed} environment ($\varrho_{rec}$) in case the $z_{phot}$ of each galaxy is used. We will regard to the \textit{True} environment as to the reference one and we will estimate how much photometric redshifts impact the density field reconstruction by comparing it to the \textit{Reconstructed} environment. It is important to keep in mind that even the definition of the \textit{True} environment is not unambiguous, as different scales and different definitions probe different physical processes.

We estimated galaxy environments as the volume density of objects inside a cylinder with base radius $R$ and height $h$, centered on each galaxy of the sample, using all other galaxies as tracers. We adopted several values of $R$ and $h$, so to have an exploration of the dependence of the environment reconstruction precision on these two parameters. In particular we varied the base radius of the cylinder between $R = 0.3$ Mpc and $R = 2$ Mpc comoving. We chose the height of the cylinder (\textit{i.e.} its length on the radial direction) proportional to redshift in two different ways, according to whether we performed the environmental estimate using the $z_{true}$ or $z_{phot}$ of galaxies. In particular $h$ is defined as 
\begin{equation}
h = \pm \varDelta z
\end{equation}

When using $z_{true}$ we adopted a $\varDelta z$ corresponding to a $dv = 1000 km/s$, through the formula
\begin{equation}\label{hdv}
\varDelta z = \frac{dv}{c}\cdot(1+z)
\end{equation}

We chose such a cylinder length to be consistent with other definitions of local environment from the literature, as this is the value which is generally adopted to estimate the density field when using spectroscopic redshifts.

In the $z_{phot}$ case, such a small height for the cylinder is useless as the errors on the redshift are much larger than that (an environmental estimate with photometric redshifts and $\varDelta z$ from equation \eqref{hdv} has been performed as well, but it will be used only as a reference for what happens when the cylinder length is very small compared to the photometric redshift uncertainty). Therefore in the $z_{phot}$ case we chose $\varDelta z$ proportional to the error on the photometric redshift
\begin{equation}\label{hn}
\varDelta z = n \cdot \sigma_{\varDelta z/(1+z)} \cdot (1+z)
\end{equation}

We varied the parameter $n$ as $n = 0.5, 1.5, 3$, so to have a total length of the cylinder $h$ ranging from the $\pm 0.5\sigma$ to $\pm 1.5\sigma$ and $\pm 3\sigma$ photometric redshift error. In order to mimick with our analysis the UltraVISTA data \citepads{2012A&A...544A.156M} we adopted a value for the photometric redshift error of $\sigma_{\varDelta z/(1+z)} = 0.01$, according to the value reported in Figure 1 of \citetads{2013A&A...556A..55I}. This Figure shows a comparison between the photometric redshifts and a sample of spectroscopic redshifts at $K_S \le 24$. We chose this value also because it is in agreement with the mean of the error values reported in Table 1 of \citetads{2013A&A...556A..55I}, weighted by the number of sources in each spectroscopic sample used to determine the error. This takes into account the fact that the spectroscopic samples used to derive the values of Table 1 of \citetads{2013A&A...556A..55I} are sometimes small, composed of a few tens of galaxies, and therefore the errors reported may not be representative of the whole spectroscopic sample at $K_S \le 24$. We will use the value of $\sigma_{\varDelta z/(1+z)} = 0.01$ to expose our main results, although we also performed several tests by choosing different values for $\sigma_{\varDelta z/(1+z)}$, which we will discuss in section \ref{photozerror}.

We calculated volume densities by dividing the number counts of objects inside each cylinder by the cylinder volume. It has to be noted that many studies rely on surface densities of galaxies inside a redshift bin to define environments instead of volume densities. We also calculated surface densities as the number counts inside each fixed aperture volume divided by the aperture base area. Nevertheless, in the following we will use volume densities to derive our results as they allow to homogenize results inside each redshift bin. In fact, our redshift bins are large enough for galaxies to have a redshift distribution inside each redshift bin. As the fixed aperture volume depends on the redshift of each galaxy, because we chose a length proportional to the photometric redshift uncertainty, galaxies closer to the lower bound of each redshift bin will have environments measured using smaller volumes compared to galaxies closer to the upper bound of each redshift bin. Therefore, even inside each redshift bin, galaxies at different redshifts will have environments measured with different volume sizes and this may bias the results, preventing a consistent comparison of environmental densities. Surface densities do not take this problem into account, because the area of each cylinder remains the same, once the aperture radius is fixed. Defining environments through the use of volume densities, instead, allows to take the cylinder dimensions into account on a galaxy by galaxy basis, in a more consistent fashion, even if it has the effect of smoothing out extreme over- and under-densities. As a test, we have also derived our results through the use of surface densities and we will discuss them in section \ref{K22sect}.

We defined high density and low-density environments using the percentiles of the volume density distributions calculated in redshift bins of width $dz = 0.25$ ranging from $z = 0$ to $z = 3$. We defined galaxies residing in environments denser than the 75th percentile of the distribution of volume densities as belonging to high-density environments, conversely we defined galaxies whose environment is less dense than the 25th percentile as belonging to low-density environments. In the following we will also refer to these environments as to $D_{75}$ and $D_{25}$ for high-density and low-density environments respectively. We also performed our analysis by chosing more extreme environments, using the 10\% and 90\% of the distribution. Nevertheless, the results found with quartiles are more stable and more significant, due to the larger statistic of the samples of galaxies costituting the various environments. Therefore, we will focus our analysis on high-density and low-density environments derived using the quartile distinction. It has to be noted that at high redshift (generally at $z > 2.5$) it may become difficult to define the 25th percentile of the volume density distribution, as the reduced sample statistics of the highest redshift bins leads the smallest volume density recovered to be shared by more than 25\% of galaxies. For this reason we decided to limit our analysis at redshifts lower than $z \le 2.5$ in order to not introduce a bias in our results. A summary of the various environmental reconstructions can be found in Table \ref{cylindervalues}.

\begin{table*}
\caption{Environmental Reconstruction. Here, $R$ and $h$ refer to the radius and the length of the volume. $dv$ and $n$ are the parameters introduced in equations \eqref{hdv} and \eqref{hn} respectively. When the parameter $n$ is used, $h$ is defined as $h = \pm n \cdot \sigma_{\varDelta z/(1+z)} \cdot (1+z)$}
\label{cylindervalues}
\centering
\begin{tabular}{c c c}
\hline\hline
Property         & \textit{True} Environment      & \textit{Reconstructed} Environment     \\
\hline
Redshift Used                & $z_{true}$           & $z_{phot}$                            \\
$R$                          & $0.3, 0.6, 1, 2$ Mpc & $0.3, 0.6, 1, 2$ Mpc                  \\
$h$                          & $dv = 1000 km/s$     & $n = 0.5, 1.5, 3$ \& $dv = 1000 km/s$ \\
$\sigma_{\varDelta z/(1+z)}$ & \ldots               & $0.003, 0.01, 0.03, 0.06$             \\
\hline
\end{tabular}
\end{table*}

\subsection{An overview of the mock data}
Before performing an accurate and quantitative analysis of the results, it is useful to have a qualitative look at the differences between the various environmental reconstructions, so to have an overall view of how the estimate is dependent on the redshift accuracy.

The three panels of Figure \ref{lightcones} show each three slices of the analysed sky field, with RA on the abscissas and redshift on the ordinate. In each panel, from left to right, the redshift on the vertical axis changes from $z_{true}$ to $z_{obs}$ and $z_{phot}$. The three panels correspond to three large redshift ranges, namely $z \in [0,1]$, $z \in [1,2]$, and $z \in [2,3]$. For the sake of clarity, the slices have been limited in declination to the central $1.2 \deg$ ($-0.6 < dec < 0.6$).

\begin{figure*}
\centering
\includegraphics[scale=0.77]{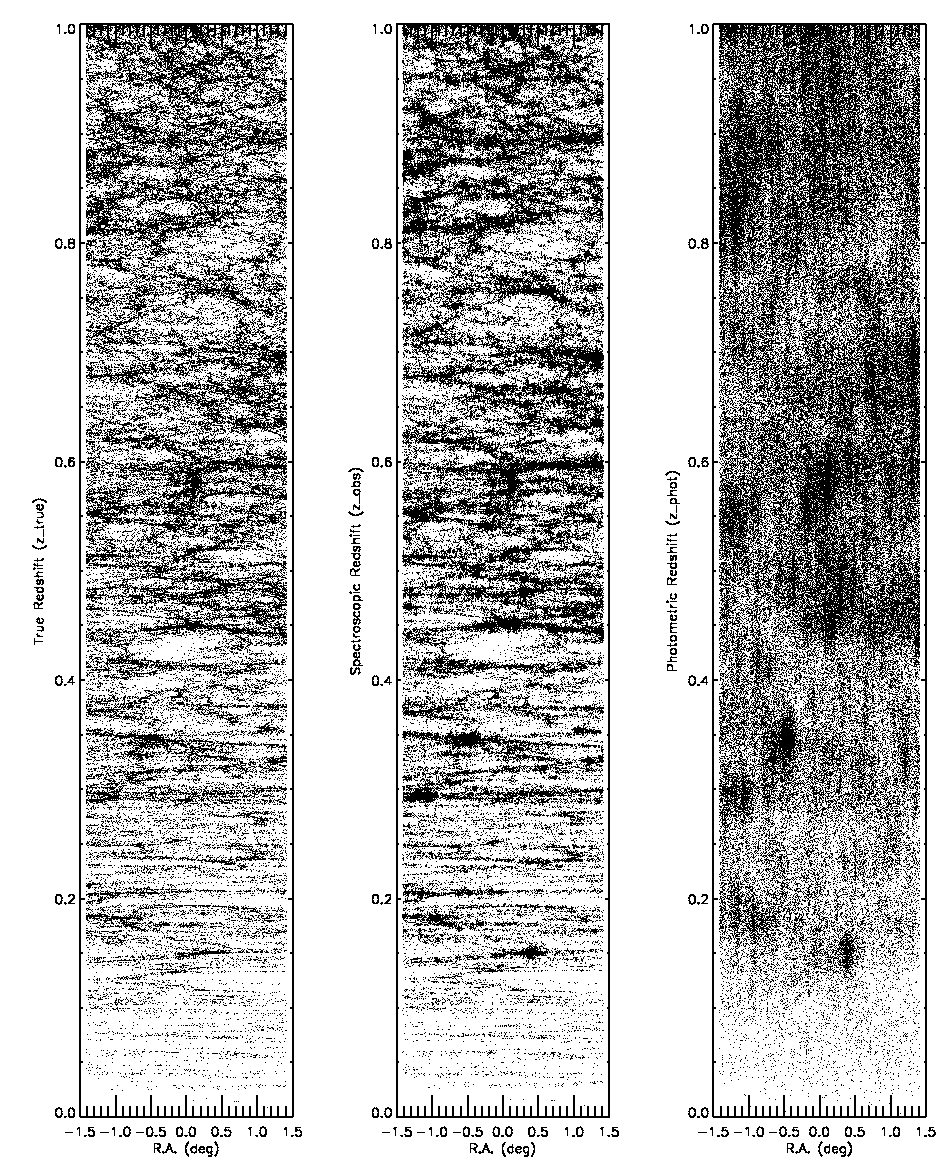}
\caption{\textit{Lightcones}. In each panel, left plot shows the lightcone obtained using the $z_{true}$ of galaxies, middle plot using $z_{obs}$ and right plot using $z_{phot}$ with $\sigma_{\varDelta z/(1+z)} = 0.01$. For the sake of clarity, the lightcones have been limited in declination to the central $1.2 \deg$. $z \in [0,1]$.}
\label{lightcones}
\end{figure*}
\begin{figure*}
\ContinuedFloat
\centering
\includegraphics[scale=0.77]{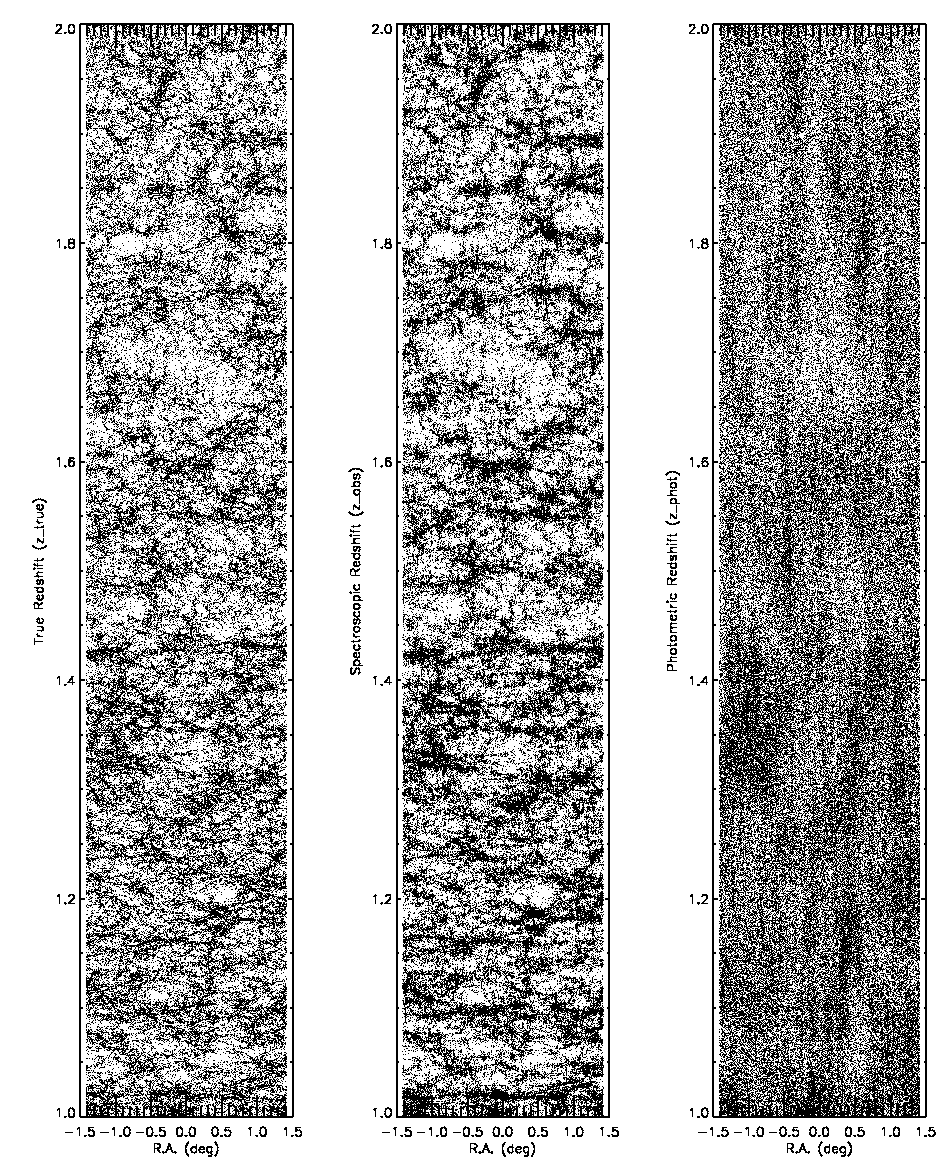}
\caption{continued, $z \in [1,2]$.}
\end{figure*}
\begin{figure*}
\centering
\ContinuedFloat
\includegraphics[scale=0.77]{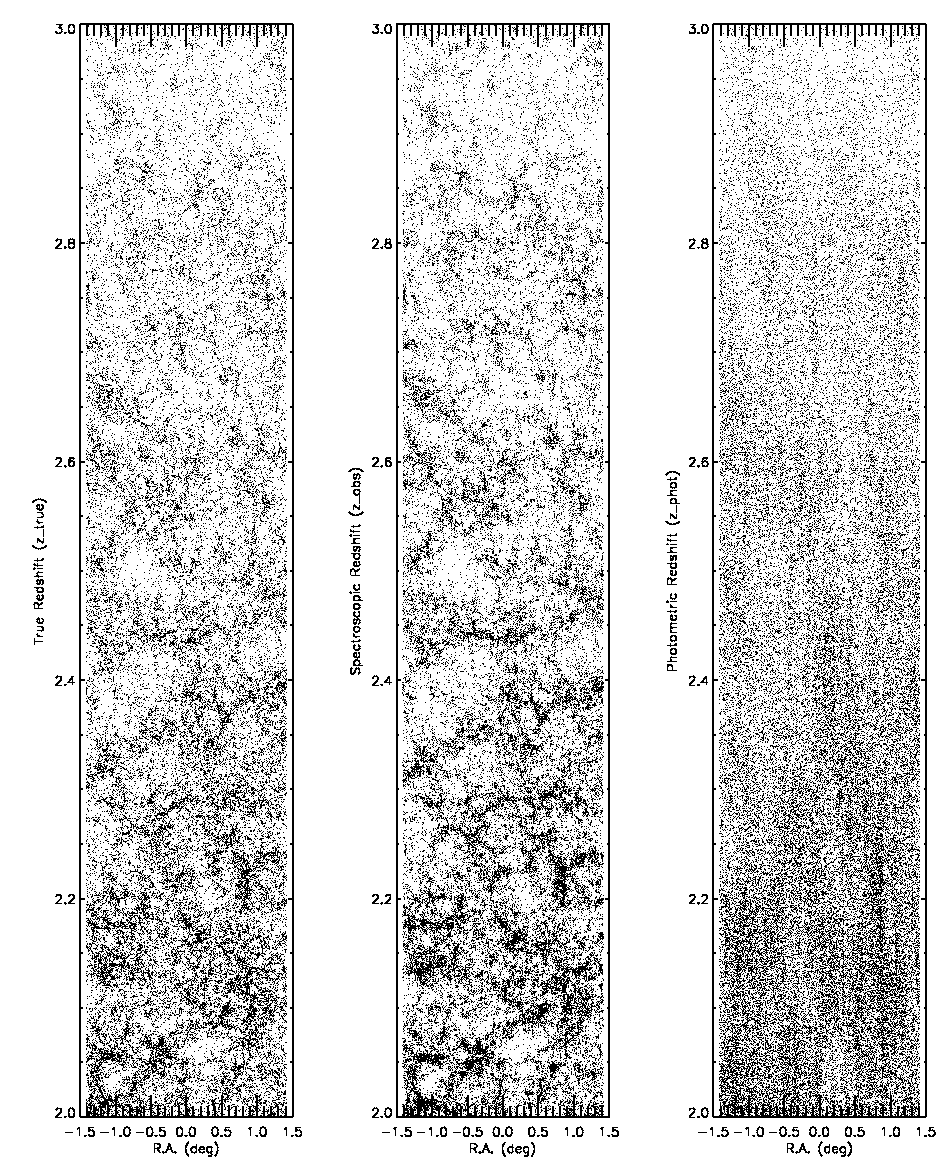}
\caption{continued, $z \in [2,3]$.}
\end{figure*}

It can be easily seen how the overdense regions that are so sharply defined in the $z_{true}$ case become more elongated when we use $z_{obs}$ and blurred when relying on the $z_{phot}$. Nevertheless, it can be seen how high-density regions are still recognisable as distinct from the mean density field and low-density areas, also in the case of $z_{phot}$ with $\sigma_{\varDelta z/(1+z)} = 0.01$. Therefore, we can still make a precise quantitative estimate of the accuracy that can be attained in reconstructing the environment for different purposes.

\begin{figure*}
\centering
\includegraphics[width=17cm]{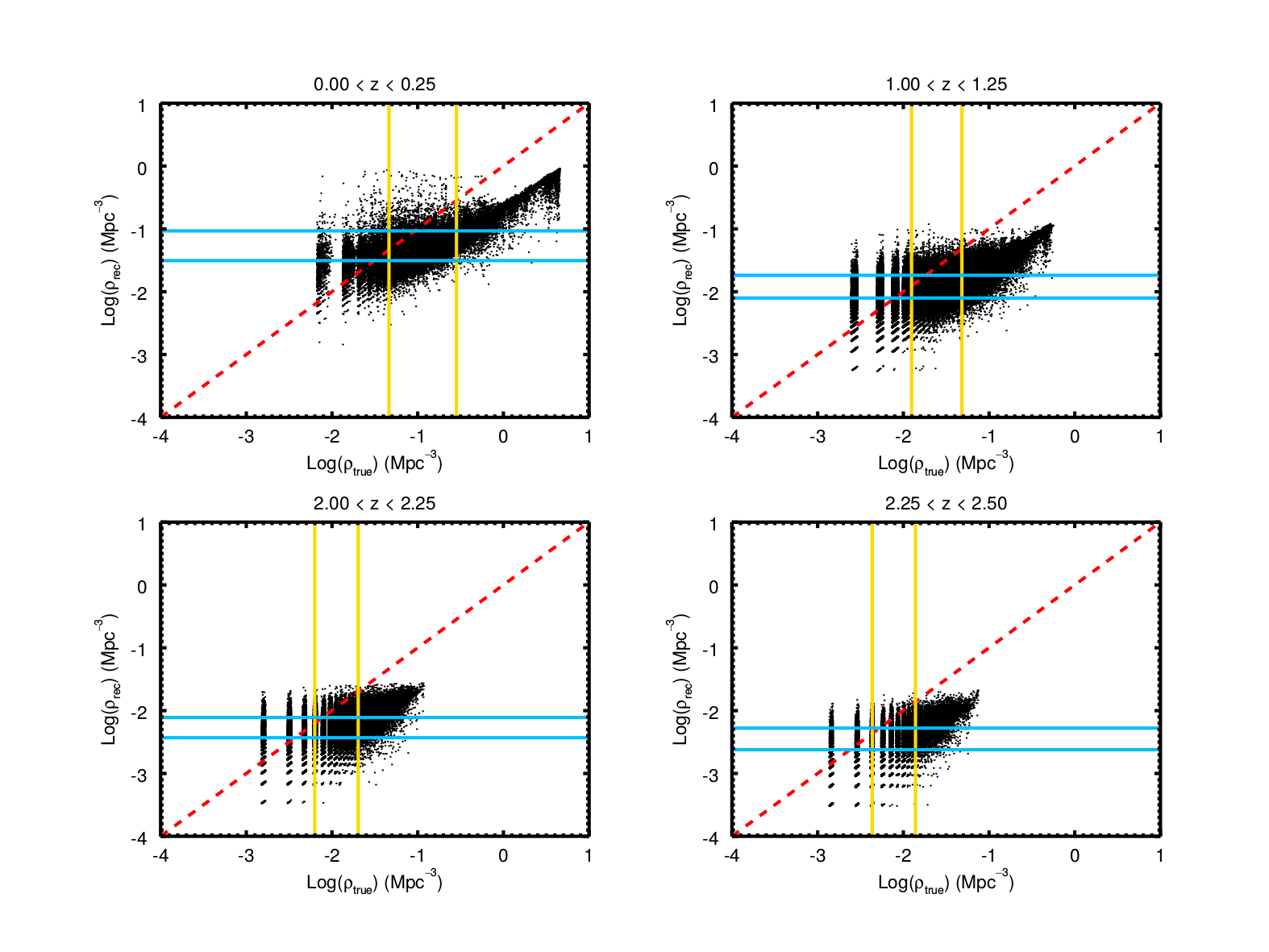}
\caption{\textit{Environment Comparison}. The figure shows the comparison between $\varrho_{true}$ and $\varrho_{rec}$ (black points) in four representative redshift bins, marked on top of each plot. Cyan horizontal lines represent the quartiles of the $\varrho_{rec}$ distribution, while yellow vertical lines are the quartiles of the $\varrho_{true}$ distribution. The 1:1 relation is reported as a red dashed line for reference. Parameters used in the environmental estimate are $R = 1$ Mpc, $n = 1.5$ and $\sigma_{\varDelta z/(1+z)} = 0.01$.}
\label{densitycomparison}
\end{figure*}

We show a first analysis of the various environmental estimates in Figure \ref{densitycomparison}. This figure shows the comparison between the \textit{Reconstructed} (with $n=1.5$) environment and the \textit{True} one on a scale of $R = 1$ Mpc. It can be seen that a correlation between the two environments is present, although the scatter is large and the points result tilted and displaced from the 1:1 relation. In fact, in each panel (which correspond to four representative redshift bins), $\varrho_{true}$ spans a wider range of volume densities (going from $0.005\div 5\: Mpc^{-3}$ at $z\sim 0$ to $0.001\div 0.1\: Mpc^{-3}$ at $z \sim 2$) compared to $\varrho_{rec}$ (which ranges from $0.005\div 1\: Mpc^{-3}$ at $z \sim 0$ to $0.002\div 0.02\: Mpc^{-3}$ at $z \sim 2$). As mentioned in Section \ref{method}, the tilt seems to be due to the fact that volume densities tend to dilute environmental variations inside each redshift bin, therefore bringing the whole environmental measurement closer to the mean value of the density field and flattening the correlation between $\varrho_{rec}$ and $\varrho_{true}$. We derived these same plots using surface densities instead of volume ones and we found a smaller tilt, with the points closer to the 1:1 relation. Nevertheless, in this work we used volume densities, as they allow to take into account the fact that, even inside the same redshift bin, galaxies at different redshifts will have environments measured with volumes of different lengths. This allows for a more self-consistent study of the density field (see also Figure \ref{SD} for a more quantitative example of the effect of surface densities on $D_{75}$ and $D_{25}$ definition).

At very low densities the distribution of the points begins to show vertical and horizontal bands, with points clustering at precise density values. This is due to the process of environmental estimate: as we count galaxies inside each cylinder and then we divide by the cylinder volume, volume densities can take only discrete values. At high densities, where the dynamic range is large, discretization effects will be less visible and the volume density distribution will become more continuous. At very low densities the effect of discretization will be more visible as there will be only a small and finite amount of galaxies inside each volume. This results in a loss of continuity in the density values in low-density environments which become progressively less discrete going towards high-densities. This discretization effect is visible also at higher density values moving at high redshifts. It seems plausible, though, that at least the most extreme environments (like those set by the quartiles of the distributions) could be well recovered. Comparisons like the one shown in Figure \ref{densitycomparison} have been studied for all the samples described later in the text (not shown).

\begin{figure*}
\centering
\includegraphics[width=17cm]{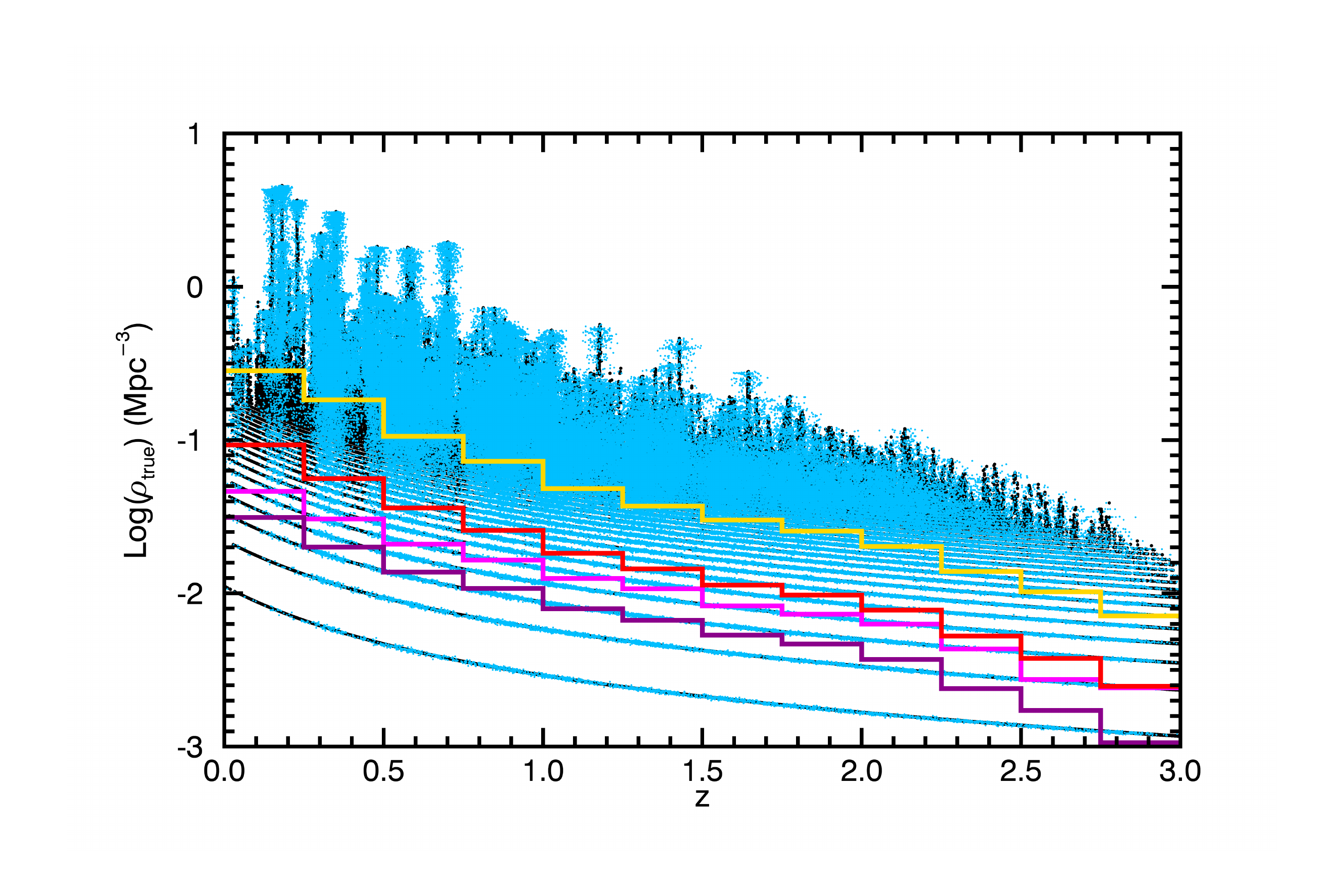}
\caption{\textit{Density-redshift relation}. This figure shows $\varrho_{true}$ as a function of redshift (black points). Cyan points highlight galaxies in high-density regions (according to $\varrho_{rec}$). The parameters used for the environmental reconstruction are $R = 1$ Mpc, $n = 1.5$ and $\sigma_{\varDelta z/(1+z)} = 0.01$. The yellow and magenta lines refer to \textit{True} environment $D_{75}$ and $D_{25}$ respectively, while the red and purple ones refer to the \textit{Reconstructed} $D_{75}$ and $D_{25}$ respectively.}
\label{densityredshift}
\end{figure*}

The two main problems that affect environment parametrization, when passing from the \textit{True} to the \textit{Reconstructed} estimate, are shown in Figure \ref{densityredshift}. This figure shows $\varrho_{true}$ as a function of redshift. Objects that are placed in high-density \textit{Reconstructed} environments, for the parameter combination $n = 1.5$, $R = 1$ Mpc and $\sigma_{\varDelta z/(1+z)} = 0.01$ are highlighted. It can be seen how high-density \textit{Reconstructed} environments are contaminated by many galaxies coming from low-density \textit{True} ones, and how $D_{75}$ environments are not fully recovered. This figure too shows the effects of discretization at low densities as a lack of continuous density values, that progressively disappears moving at high-density values. We will study all these effects in detail in the following sections.

\subsection{The definition of Recovery and Contamination fractions}
\label{recncon}
The assessment of how close $\varrho_{rec}$ is to $\varrho_{true}$ is no simple matter. Several parameters play a role in determining the accuracy of the density field reconstruction. An exploration of the parameter space is needed in order to study the degeneracies between $R$ and $\varDelta z$ of the cylinders and the effect induced by chosing redshifts with worsening precision. In particular, in the following analysis we looked for two distinct effects in the environmental reconstruction, which we will call \textquotedblleft Recovery\textquotedblright$\:$ and \textquotedblleft Contamination\textquotedblright.

With Recovery ($f_{Rec}$) we indicate the fraction of galaxies that are correctly placed in either high-density or low-density regions according both to $\varrho_{true}$ and $\varrho_{rec}$. In particular, $N_{High}^{True}$ and $N_{High}^{Rec}$ are the number of galaxies in high-density environments according to $\varrho_{true}$ and $\varrho_{rec}$ respectively (and correspondingly $N_{Low}^{Rec}$ and $N_{Low}^{True}$ for low-density environments). So, if $N_{HH}$ is the number of galaxies that are placed in a high-density environment according both to $\varrho_{true}$ and $\varrho_{rec}$, and correspondingly $N_{LL}$ for low-density environments, then the Recovery fraction is defined as
\begin{equation}
f_{Rec} = 
\begin{cases}
\frac{N_{HH}}{N_{High}^{True}} & \text{for $D_{75}$}\\
\frac{N_{LL}}{N_{Low}^{True}} & \text{for $D_{25}$}\\
\end{cases}
\end{equation}

Therefore, a $f_{Rec}$ of 1 means that all the galaxies that are in the high-density (or low-density) \textit{Reconstructed} environments are placed in the correct \textit{True} density quartile.

It is then useful to calculate the Contamination fraction, $f_{Con}$. This quantity is the fraction of galaxies that are placed in a density quartile according to $\varrho_{rec}$ that actually come from the opposite quartile according to $\varrho_{true}$. If $N_{HL}$ is the number of galaxies that are placed in a high-density environment when relying on $\varrho_{rec}$, but that actually come from a low-density environment when relying on $\varrho_{true}$ (and conversely $N_{LH}$ is the number of galaxies that are placed in a low-density environment when relying on $\varrho_{rec}$, but that actually come from a high-density environment when relying on $\varrho_{true}$), then the Contamination fraction ($f_{Con}$) is expressed as
\begin{equation}
f_{Con} =
\begin{cases}
\frac{N_{HL}}{N^{Rec}_{High}} & \text{for $D_{75}$}\\
\frac{N_{LH}}{N^{Rec}_{Low}} & \text{for $D_{25}$}\\
\end{cases}
\end{equation}

\begin{figure*}
\centering
\includegraphics[width=17cm]{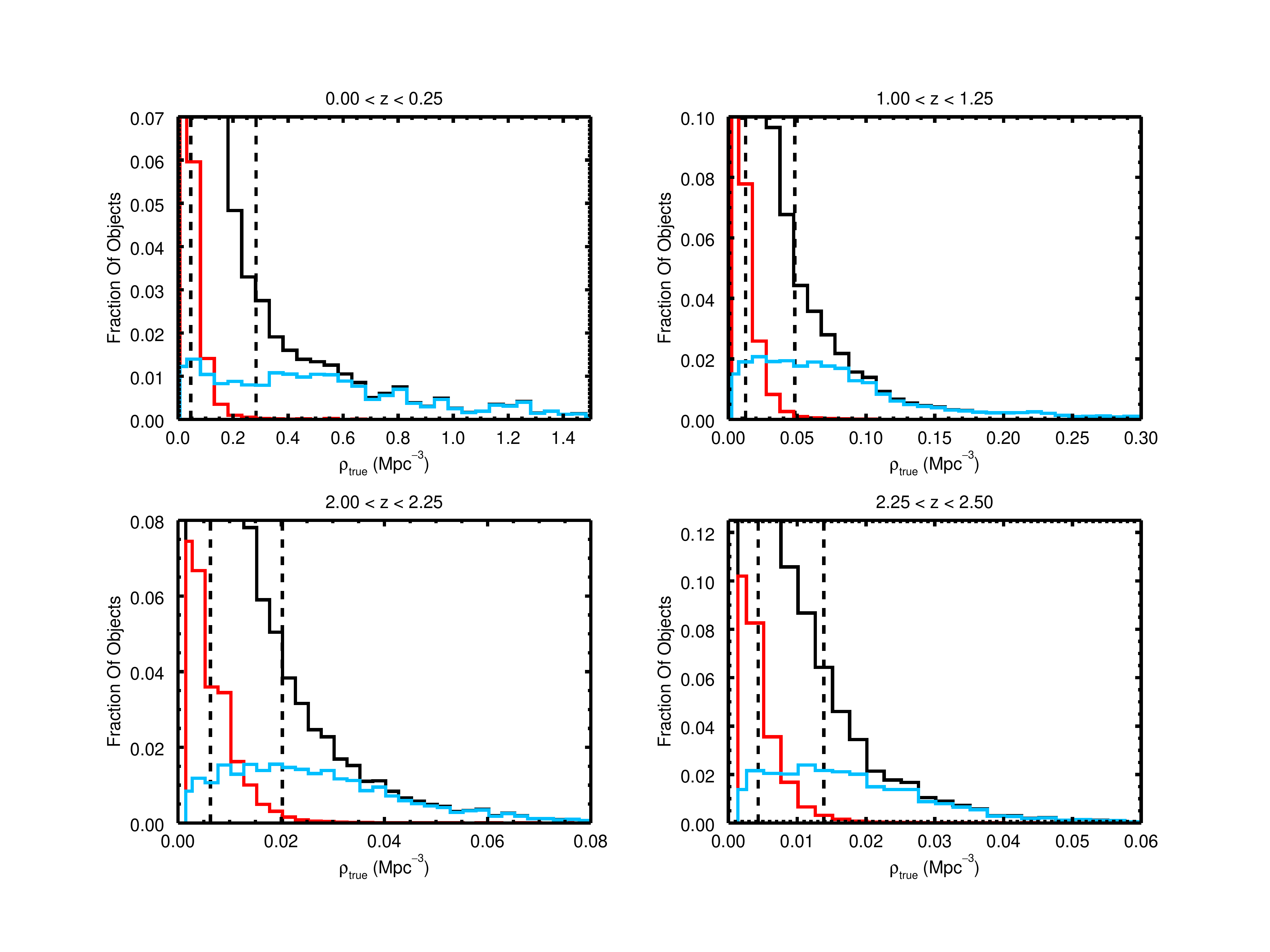}
\caption{\textit{Distributions Comparison}. This figure shows the distribution in terms of $\varrho_{true}$ of the high-density and low-density samples, identified using $\varrho_{rec}$ in four representative redshift bins (marked on top of each plot). In particular, the black histogram is the $\varrho_{true}$ distribution shown as a reference, the cyan histogram refers to $D_{75}$ $\varrho_{rec}$, while the red histogram refers to $D_{25}$ $\varrho_{rec}$. The environmental estimate has been performed with $n = 1.5$, $R = 1$ Mpc and $\sigma_{\varDelta z/(1+z)} = 0.01$. The two vertical dashed lines are the values of environmental density corresponding to the 25th and 75th percentile of the $\varrho_{true}$ distribution.}
\label{distributioncomparisonsigma}
\end{figure*}

We expect both these quantities to vary with redshift as the reconstruction of the environment is more difficult for high-redshift galaxies due to the lower accuracy of the photometric redshift estimate. A good way to visualize these quantities is shown in Figure \ref{distributioncomparisonsigma}. We report only four redshift bins for reference.

In this figure we show how the high-density sample (objects above the 75th percentile) and the low-density one (objects below the 25th percentile) identified using $\varrho_{rec}$, with $n = 1.5$, $R = 1$ Mpc and $\sigma_{\varDelta z/(1+z)} = 0.01$, are distributed according to $\varrho_{true}$ (whose distribution is also reported for reference, together with vertical dashed lines corresponding to 25\% and 75\% of the $\varrho_{true}$ distribution). It can be seen that at very high densities and very low ones  the $D_{75}$ and $D_{25}$ $\varrho_{rec}$ distributions follow closely the corresponding parts of the $\varrho_{true}$ distribution. The better a \textit{Reconstructed} distribution follows the \textit{True} one, the higher its Recovery fraction $f_{Rec}$. The ideal case of perfect reconstruction would imply that the \textit{Reconstructed} distribution of the high-density environments followed the $\varrho_{true}$ distribution down to the line of the 75th percentile and then dropped to zero, or that the low-density \textit{Reconstructed} distribution rose following the $\varrho_{true}$ one up to the 25th percentile line and then dropped to zero as well.

However, it can be seen how the distributions of the $\varrho_{rec}$ $D_{25}$ and $D_{75}$ environments have tails extending to $\varrho_{true}$ values of the opposite quartile. This means that a fraction of objects identified as high- or low-density ones according to $\varrho_{rec}$ actually comes from low- or high-density regions according to $\varrho_{true}$. This fraction contributes to the Contamination fraction of the sample ($f_{Con}$).

\section{Results for the best case of $\sigma_{\varDelta z/(1+z)} = 0.01$}
\label{results}
As different physical processes operate on different scales, it is of great importance to determine what environmental scale can be better reconstructed with a fixed aperture volume of given $R$ and $\varDelta z$. We propose a solution to this problem by investigating wich is the best combination of $R$ and $\varDelta z$ to optimize the fixed aperture volume for reconstructing a given physical scale. This way of analysing the problem also accounts for different issues in the creation of the \textit{Reconstructed} environment that can potentially lead to biases in the conclusions.

First of all, also when relying on mock catalogues such as we do in this work, there is no unambiguous way to define a reference \textit{True} environment. There is no reason for which the environment estimated with a fixed aperture (or even a nearest neighbour) method should completely describe the true spatial distribution of galaxies, even when relying on $z_{true}$ with no intrinsic error.

Secondly, different spatial scales traced by different fixed aperture radii probe different physical mechanisms \citepads[see \emph{e.g.} figure 10 of][]{2003ApJ...591...53T}: smaller apertures (in our case $R = 0.3 \div 0.6$ Mpc) will more likely trace interacting pairs, very small groups or the very centre of bound structures, whereas greater spatial scales (as in this work the $R = 2$ Mpc case) will be more sensitive to larger structures and clusters. Therefore, finding the best way to study environments on various scales will help to create a better consistent picture of the role of the environment in galaxy evolution.

\subsection{The differential effect of $n$ and $R$ on environmental reconstruction}
\label{rndiffeffect}
\begin{figure}
\resizebox{\hsize}{!}{\includegraphics{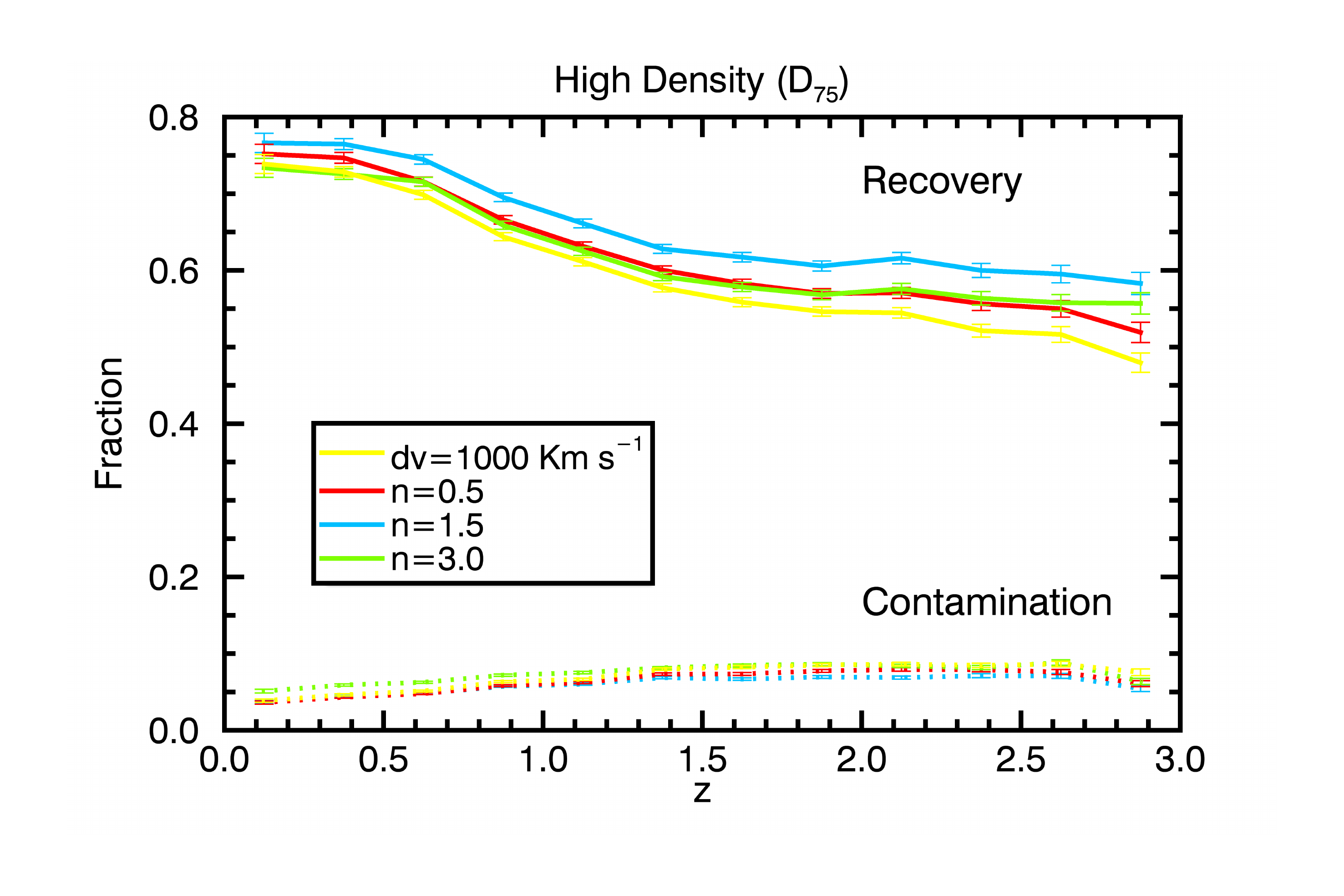}}
\resizebox{\hsize}{!}{\includegraphics{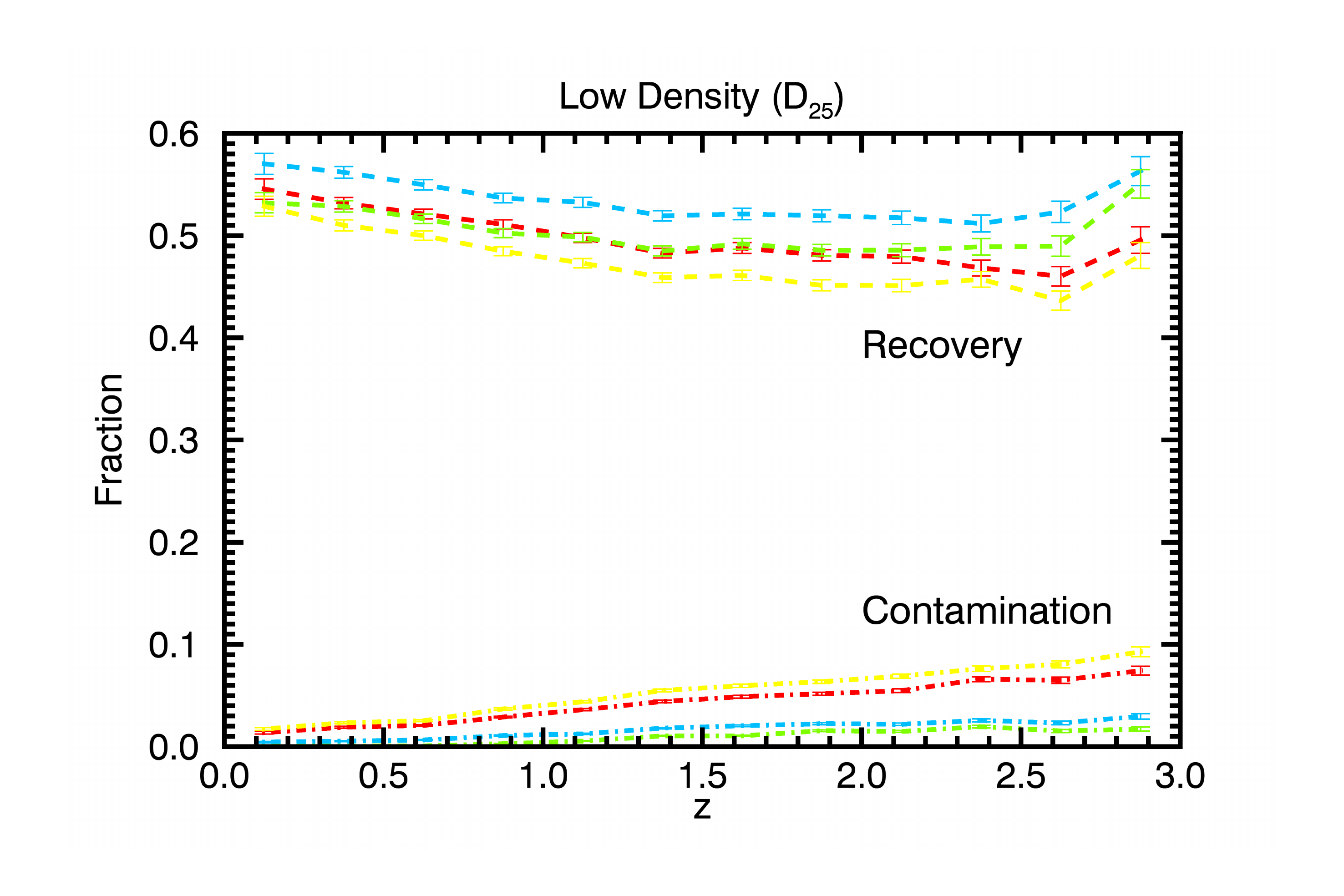}}
\caption{\textit{Recovery and Contamination fractions - varying $n$}. This figure shows $f_{Rec}$ (solid and dashed lines) and $f_{Con}$ (dotted and dot-dashed lines) as a function of redshift. Bottom panel refers to low-density environments while top panel refers to high-density environments. The various curves are color-coded according to the various values of $n$ (yellow: $dv = 1000 km/s$, red: $n = 0.5$, cyan: $n = 1.5$, and green: $n = 3$). The radius has been fixed to $R_{T} = R_{R} = 1$ Mpc and the uncertainty on the photometric redshift is $\sigma_{\varDelta z/(1+z)} = 0.01$. Note the different scale on the ordinate axis between high-density and low-density environments.}
\label{fractionssigma}
\end{figure}

We performed several tests on the data in order to explore how the $R$ and $n$ parameters used in the fixed aperture volume definition affect the reconstruction of the density field. Figure \ref{fractionssigma} shows the effect of varying the $n$ value on the environmental reconstruction. It can be seen that for all values of $n$ the reconstruction of high-density environments is fairly accurate and anyway better than that of low-density environments. The Recovery fraction in the high-density case is above 55\% up to $z \sim 2$, and above 60\% at $z < 1$. Contamination fractions are always below 10\% at all redshifts, although we do not extend our analysis farther than $z \sim 2.5$, as the reduced sample size in the farthest redshift bins may have a predominant role in creating the trends observed in the data.

The situation is slightly worse for low-density environments, which are reconstructed in a less precise way, due to the fact that low number counts will have a higher error. Contamination from high-density interlopers is low, but $f_{Rec}$ is never above 60\%.

The most important feature that can be observed in this figure is, nevertheless, the fact that the length of the cylinder used for environmental reconstruction has indeed an effect on how accurately the environment is recovered. In fact, both volume heights that are too small (such as $n = 0.5$ or $dv = 1000 km/s$) or too large (such as $n = 3$) compared to the $3\sigma$ error of the photometric redshifts have the effect of worsening the reconstruction of the environment, increasing $f_{Con}$ and decreasing $f_{Rec}$. To summarize, we found that when dealing with high-precision photometric redshifts ($\sigma_{\varDelta z/(1+z)} = 0.01$) a fixed aperture volume with a length roughly of the scale of $\pm 1.5 \sigma$ error around $z_{phot}$ is the one that grants the best environmental reconstruction. 

\begin{figure}
\resizebox{\hsize}{!}{\includegraphics{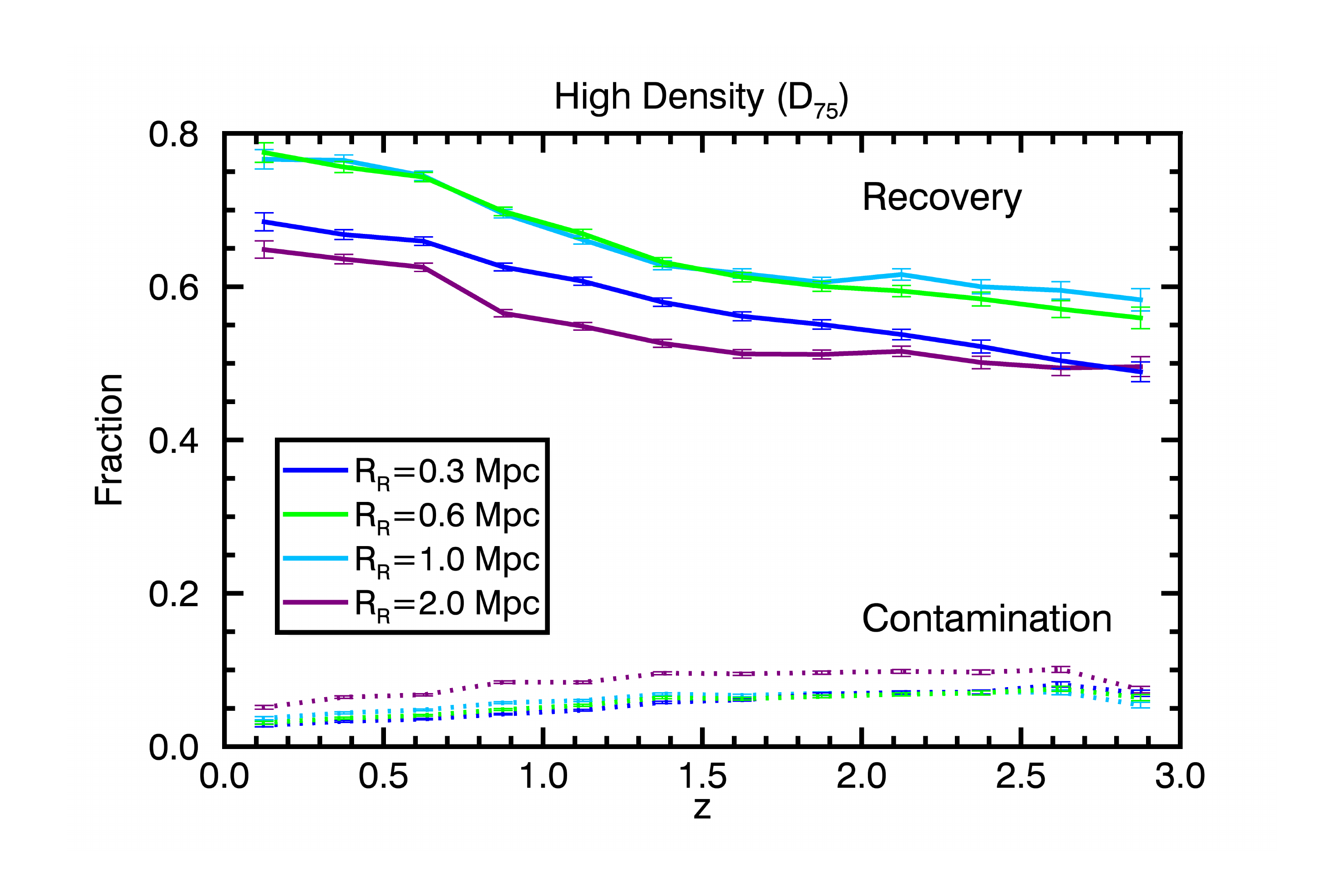}}
\resizebox{\hsize}{!}{\includegraphics{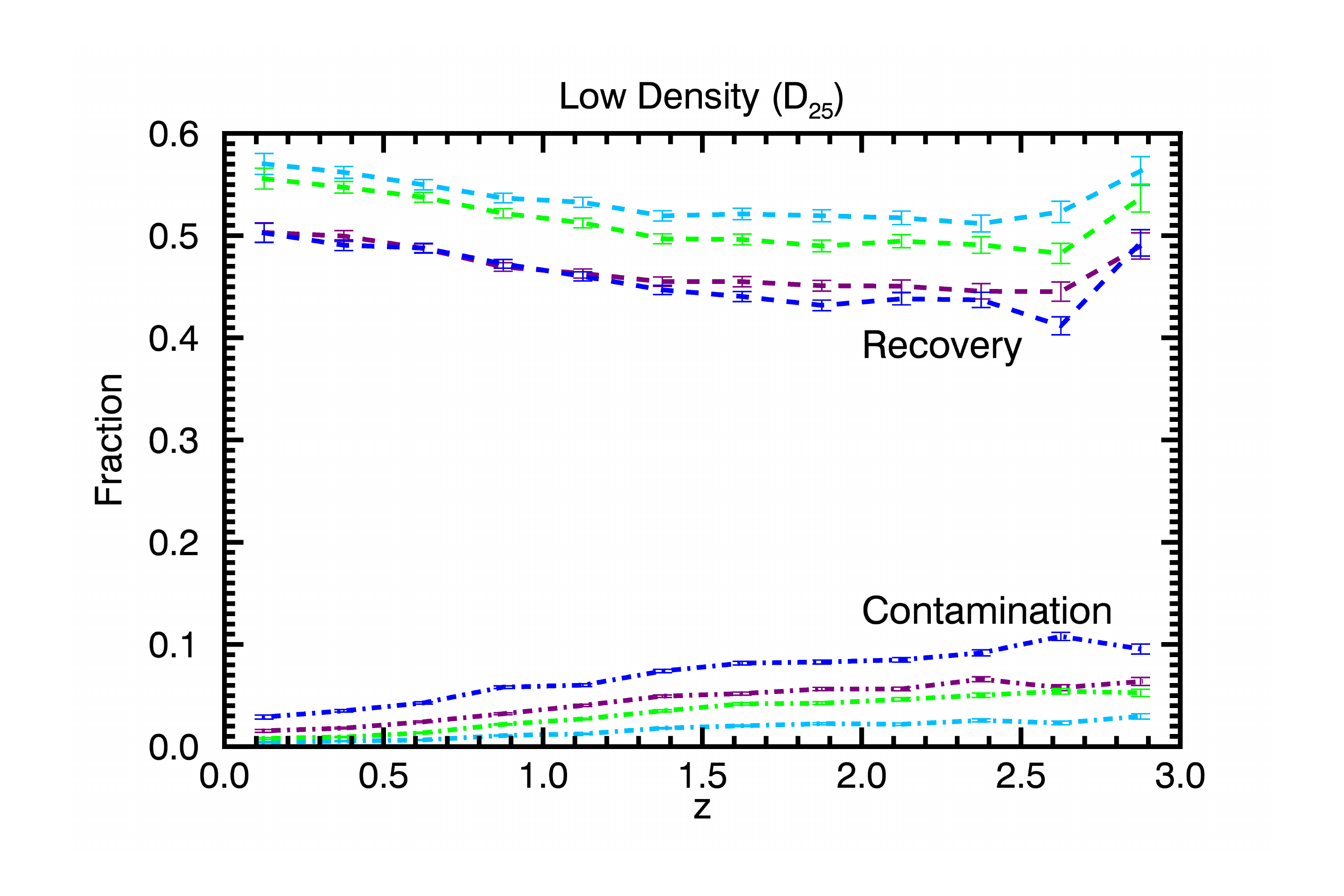}}
\caption{\textit{Recovery and Contamination fractions - varying $R_{R}$}. This figure shows $f_{Rec}$ (solid and dashed lines) and $f_{Con}$ (dotted and dot-dashed lines) as a function of redshift. Bottom panel refers to low-density environments while top panel refers to high-density environments. The various curves are color-coded according to the various values of $R_{R}$ (blue: $R_{R} = 0.3$ Mpc, green: $R_{R} = 0.6$ Mpc, cyan: $R_{R} = 1$ Mpc, and purple: $R_{R} = 2$ Mpc). The volume height has been fixed to $n = 1.5$, the uncertainty on the photometric redshift is $\sigma_{\varDelta z/(1+z)} = 0.01$ and the \textit{True} environment has been estimated on a scale $R_{T} = 1$ Mpc. Note the different scale on the ordinate axis between high-density and low-density environments.}
\label{fractionsradius}
\end{figure}

We then fixed the height of the volume used, in order to check the effect that a varying fixed aperture radius from $R = 0.3$ Mpc to $R = 2$ Mpc may have on the process. For clarity, in the following we will distinguish between $R_{T}$ when we refer to the radius used to estimate $\varrho_{true}$ and $R_{R}$ when we refer to the radius used to estimate $\varrho_{rec}$. Results are shown in Figure \ref{fractionsradius}. Again it can be seen that the \textit{Reconstructed} environment is not too different from the \textit{True} one, with $f_{Rec}$ always above 55\% up to redshift $z \sim 2$ (above 60\% at redshift $z \lesssim 1$) and $f_{Con}$ always below 10\% at all redshifts. Again, the environmental reconstruction is better for high-density environments than for low-density ones.

Figure \ref{fractionsradius} shows that also the aperture radius has an effect on the environmental reconstruction when dealing with high-precision photometric redshifts ($\sigma_{\varDelta z/(1+z)} = 0.01$). In particular both too large radii (\textit{e.g.} $R_{R} = 2$ Mpc) and too small ones (\textit{e.g.} $R_{R} = 0.3$ Mpc) compared to $R_{T}$ (in this case $R_{T} = 1$ Mpc) have the effect of lowering the accuracy of the \textit{Reconstructed} environment, increasing $f_{Con}$ and decreasing $f_{Rec}$. Thus, it is possible to conclude that increasing or decreasing too much the fixed aperture size has the effect of worsening the precision of the environmental reconstruction. To summarize, we found that when dealing with high-precision photometric redshifts ($\sigma_{\varDelta z/(1+z)} = 0.01$) a value of $R_{R} \simeq R_{T}$ is the one that optimizes the environmental reconstruction. These results are generalized for all aperture radii and various values of $\sigma_{\varDelta z/(1+z)}$ in section \ref{photozerror}.

\subsection{Additional tests on the survey magnitude limit and on the use of surface densities}
\label{K22sect}
We also performed additional tests on the samples, in order to assess the effect of other issues on the environmental reconstruction. In particular, in this section we discuss the effect of restricting the sample used in the environmental definition to only bright objects at $K \le 22$ and we propose an example of the effect introduced by relying on surface densities for the definition of high- and low-density environments. 

We performed the analysis explained so far using only the brightest galaxies, both as targets for the estimate of the environment and as tracers of the density field. This is a preliminary step to simulate the case of shallower surveys, as many large-area photometric redshift sky surveys are not able to reach K-band magnitude values as deep as $K = 24$. We therefore reduced the sample to 230\,050 objects, through a cut to magnitude $K \le 22$ and we used this sample to estimate both $\varrho_{true}$ and $\varrho_{rec}$. We found that all the results exposed so far are maintained, both as a function of $n$ and $R$ only out to $z \sim 1.5$.

\begin{figure}
\resizebox{\hsize}{!}{\includegraphics{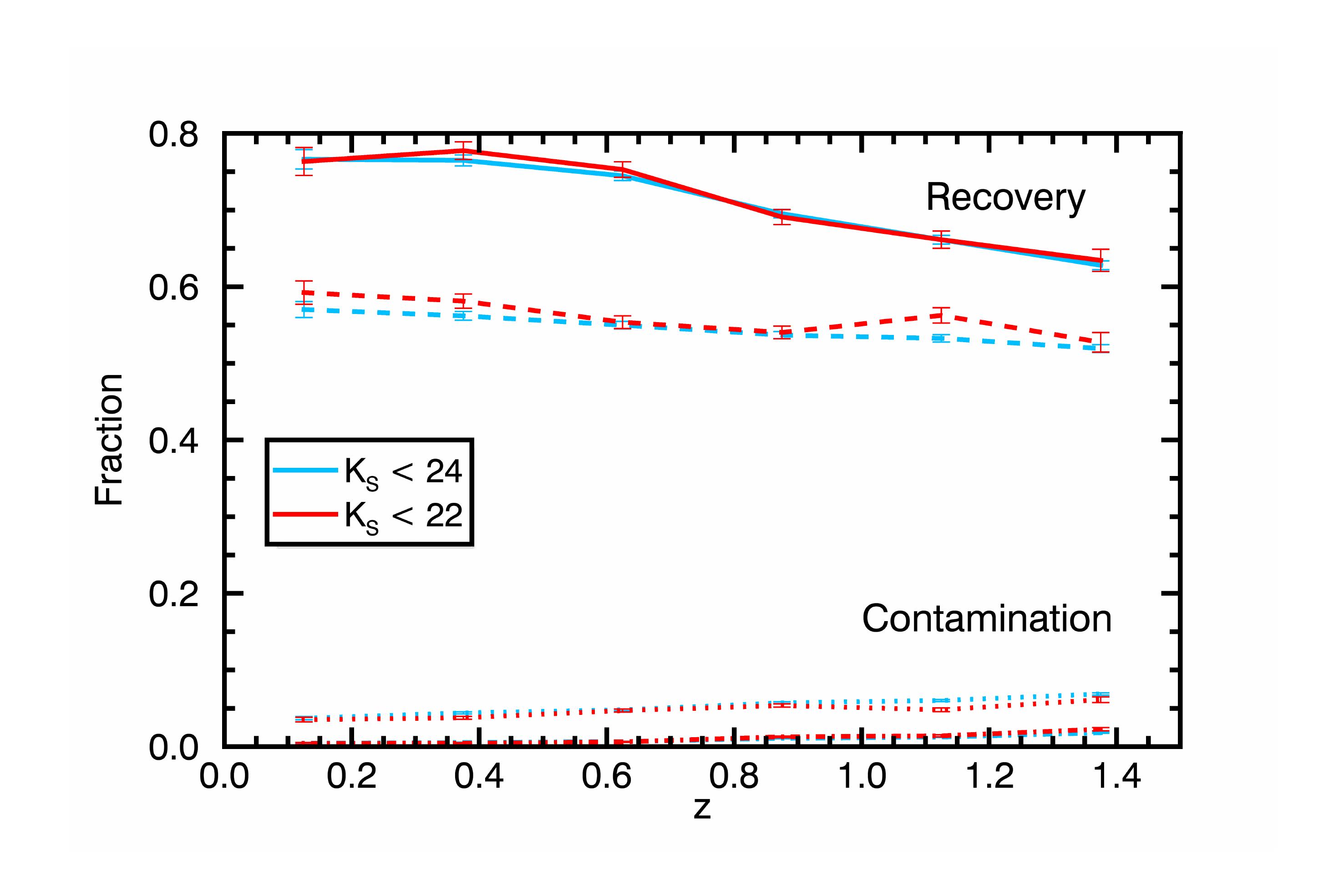}}
\caption{\textit{Recovery and Contamination fractions - $K \le 22$}. This figure shows $f_{Rec}$ (solid and dashed lines) and $f_{Con}$ (dotted and dot-dashed lines) as a function of redshift. Solid and dotted lines are for high-density environments, while dashed and dot-dashed lines are for low-density environments. Red curves are for $K \le 22$ objects, while cyan 
curves are for $K \le 24$ objects. The volume height has been fixed to $n = 1.5$, the uncertainty on the photometric redshift is $\sigma_{\varDelta z/(1+z)} = 0.01$ and both $\varrho_{true}$ and $\varrho_{rec}$ have been estimated on a scale $R_{T} = R_{R} = 1$ Mpc.}
\label{K22}
\end{figure}

We show an example of this in Figure \ref{K22}, where we report $f_{Rec}$ and $f_{Con}$ for both $D_{75}$ and $D_{25}$ as a function of redshift. Recovery and Contamination fractions from the $K \le 24$ case of Figure \ref{fractionssigma} are reported for reference. It can be seen how the fraction values in the case of $K \le 22$ follow closely those of the $K \le 24$, being between $60\% \div 80\%$ (Recovery) and lower than $10\%$ (Contamination) out to $z \sim 1.5$. We could not perform this analysis up to redshifts higher than $z = 1.5$ as the statistics in each redshift bin becomes too small to correctly define high- and low-density environments. At these redshifts (not shown here) a comparison becomes therefore impossible between the $K \le 24$ and the $K \le 22$ case.

It can be concluded that an estimate of the density field based only on the brightest objects is accurate enough to reproduce the trends observed for fainter K-band magnitude values, but in a narrower redshift range. This test shows how an environmental estimate performed with only a population of easily-observable bright $K \le 22$ objects is accurate, but needs either a large survey area or an extension to fainter maginitudes (such as $K \le 24$) in order to increase the sample statistics enough to draw conclusions for high-redshift objects. Nevertheless, this analysis shows that the fixed aperture method that we adopted is robust enough not to depend too much on the magnitude cut of the sample. The limitation in such an approach is, nevertheless, the fact that the luminous $K \le 22$ tracers are too few to extend the analysis beyond $z = 1.5$.

\begin{figure}
\resizebox{\hsize}{!}{\includegraphics{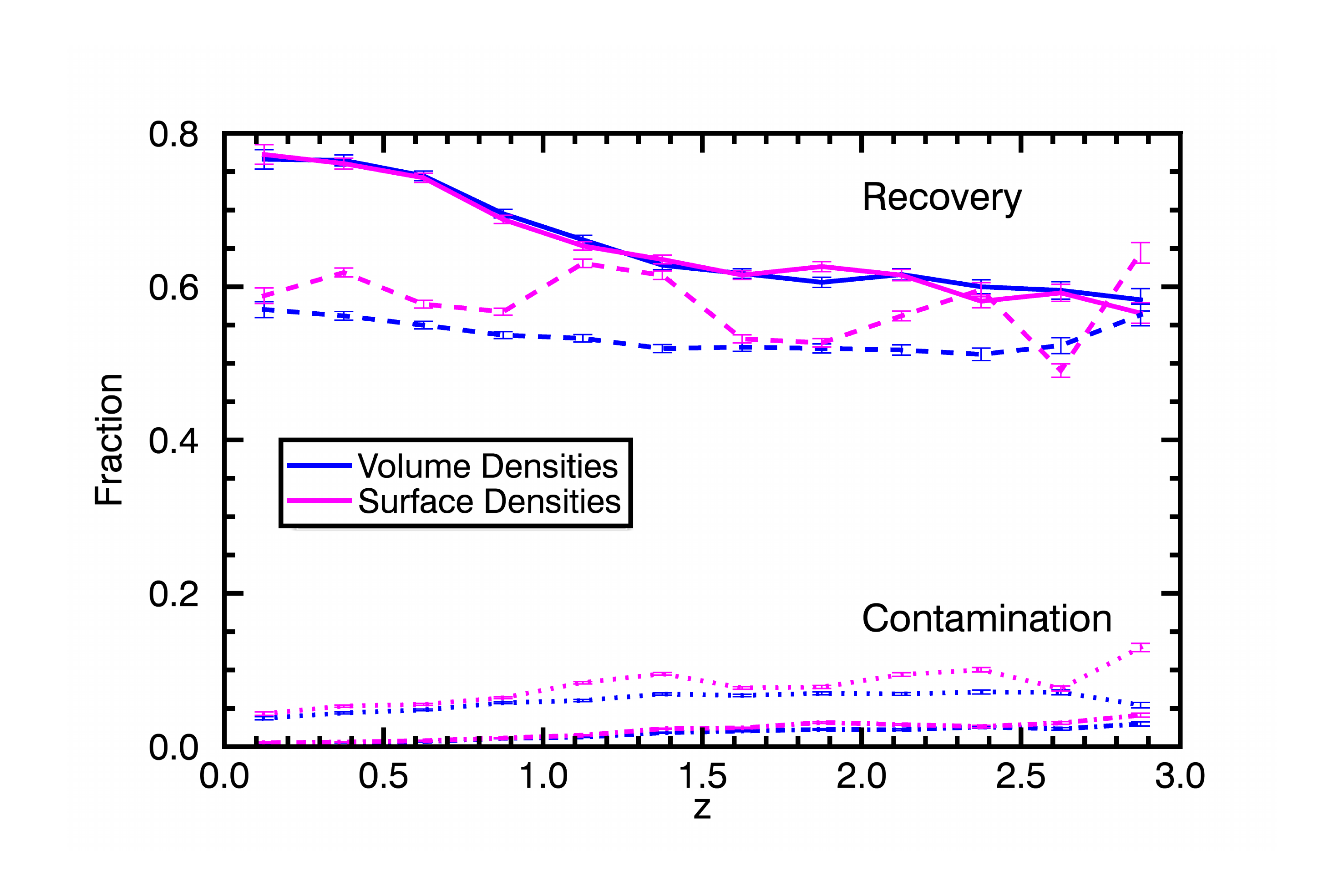}}
\caption{\textit{Recovery and Contamination fractions - Surface Densities}. This figure shows $f_{Rec}$ (solid and dashed lines) and $f_{Con}$ (dotted and dot-dashed lines) as a function of redshift. Solid and dotted lines are for high-density environments, while dashed and dot-dashed lines are for low-density environments. Magenta curves are for environments defined through the use of surface densities, while blue curves are for environments defined through the use of volume densities. The volume height has been fixed to $n = 1.5$, the uncertainty on the photometric redshift is $\sigma_{\varDelta z/(1+z)} = 0.01$ and both $\varrho_{true}$ and $\varrho_{rec}$ have been estimated on a scale $R_{T} = R_{R} = 1$ Mpc.}
\label{SD}
\end{figure}

Figure \ref{SD}, instead, shows $f_{Rec}$ and $f_{Con}$ for high-density and low-density environments, when volume densities and surface densities are used for the environmental definition. It can be seen that, although in the high-density case the agreement between volume and surface densities is remarkable, in low-density environments Recovery fractions defined through the use of surface densities are more noisy and tend to oscillate around the values set by Recovery fractions in the volume density case. The peaks in the $f_{Rec}$ values of the $D_{25}$ galaxies are caused by the discretization effect already described in section \ref{method}, which is enhanced by the use of surface densities compared to the volume density case for the \textit{True} environment. This represents a problem in the definition of the low-density environments, as even for low redshifts the value of the surface density distribution corresponding to the 25th percentile will be shared by more than 25\% of galaxies. Therefore, as the discretization effect is lower for the \textit{Reconstructed} environment case, the Recovery fraction will be higher. Nevertheless, this must be regarded as a spurious effect and as a signal that volume densities are preferrable for the definition of high-density and low-density environments. Contamination fractions show the same effect and are slightly larger than in the volume density case, being around 15\% at $z > 1$. As stated in the previous sections, volume densities grant a more accurate environmental reconstruction as they allow to account for differences in the cylinder length used for environmental definition on a galaxy by galaxy basis.

\section{The importance of $\sigma_{\varDelta z/(1+z)}$}
\label{photozerror}
So far, we have chosen a value of $\sigma_{\varDelta z/(1+z)} = 0.01$ (see section \ref{method}). In the overview of the photometric redshift surveys this is a fairly optimistic value, as very few surveys can reach this kind of precision (\textit{e.g.} the expected value of the photometric redshift precision for the Euclid survey is $\sigma_{\varDelta z/(1+z)} = 0.03 \div 0.05$). Our choice for such a small value has been determined by the fact that this is the value of the uncertainty of photometric redshifts in the COSMOS-UltraVISTA Survey sample (see \citeads{2012A&A...544A.156M}, \citeads{2013A&A...556A..55I}) to which we plan to apply the results of this work. Nevertheless, many other surveys show a variety of photometric redshift errors and it is possible that the results exposed in the previous sections do not hold when the photometric redshift error is larger.

To test this possibility we have redone the analysis, exploring several values for the photometric redshift error. In particular, we have chosen to vary $\sigma_{\varDelta z/(1+z)}$ from $\sigma_{\varDelta z/(1+z)} = 0.003$ to $\sigma_{\varDelta z/(1+z)} = 0.03, 0.06$. We chose these values as representative of various future and ongoing surveys, in particular $\sigma_{\varDelta z/(1+z)} = 0.003$ is the value expected for the Javalambre Physics of the accelerating universe Astronomical Survey\footnote{\url{http://www.j-pas.org/}} (J-PAS, PI: Benitez, \citeads[see][]{2014arXiv1403.5237B}), $\sigma_{\varDelta z/(1+z)} = 0.03$ is the minimum error expected for the Euclid Survey\footnote{\url{http://www.euclid-ec.org/}} (PI: Mellier, \citeads[see][]{2011arXiv1110.3193L}), $\sigma_{\varDelta z/(1+z)} = 0.06$ is the error on photometric redshifts derived for the sources in previous releases of the COSMOS Survey\footnote{\url{http://www.cosmos.astro.caltech.edu/}} (PI: Scoville, see \citeads{2007ApJS..172....1S}, \citeads{2007ApJS..172...99C}, \citeads{2009ApJ...690.1236I}) and used for example in \citeads{2015A&A...576A.101M}.

\begin{figure}
\resizebox{\hsize}{!}{\includegraphics{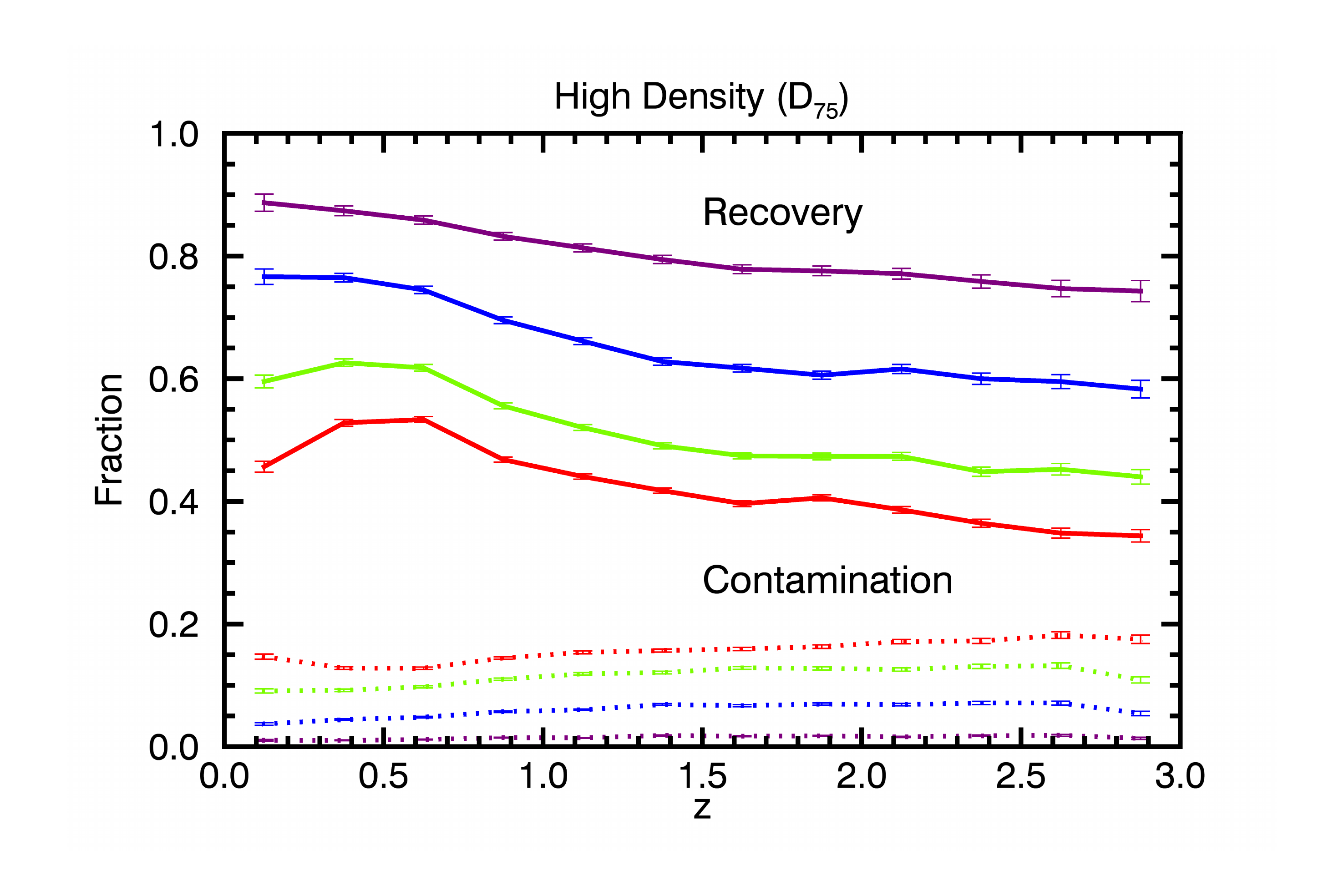}}
\resizebox{\hsize}{!}{\includegraphics{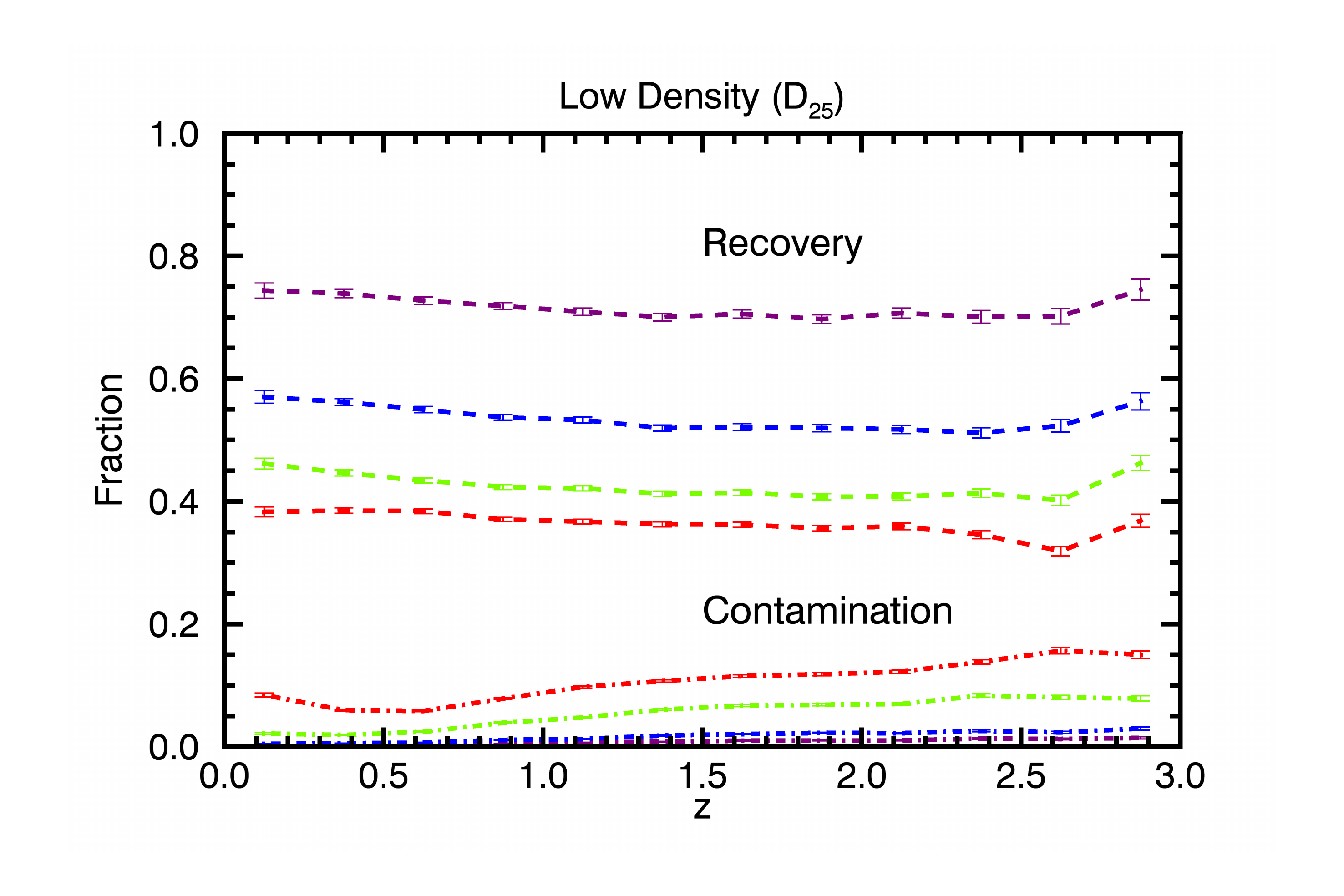}}
\caption{\textit{Recovery and Contamination fractions - varying $\sigma_{\varDelta z/(1+z)}$}. This figure shows $f_{Rec}$ (solid and dashed lines) and $f_{Con}$ (dotted and dot-dashed lines) as a function of redshift. Bottom panel refers to low-density environments while top panel refers to high-density environments. The various curves are color-coded according to the 
various values of $\sigma_{\varDelta z/(1+z)}$ (purple: $\sigma_{\varDelta z/(1+z)} = 0.003$, blue: $\sigma_{\varDelta z/(1+z)} = 0.01$, green: $\sigma_{\varDelta z/(1+z)} = 0.03$, and red: $\sigma_{\varDelta z/(1+z)} = 0.06$). The volume parameters for environmental reconstruction have been kept fixed to $R_{R} = R_{T} = 1$ Mpc and $n = 1.5$.}
\label{allsurvey}
\end{figure}

A first, expected result is that the environmental reconstruction is more difficult when using photometric redshifts with large uncertainties. Figure \ref{allsurvey} shows how both in the high-density and low-density situations $f_{Rec}$ decreases and $f_{Con}$ increases with increasing $\sigma_{\varDelta z/(1+z)}$.

In particular, for high-density environments $f_{Rec}$ is always above 75\% (and close to 90\% at $z \sim 0$) in the high-accuracy $\sigma_{\varDelta z/(1+z)} = 0.003$ case, and progressively decreases to values slightly above 40\% at all redshifts in the low accuracy $\sigma_{\varDelta z/(1+z)} = 0.06$ one. Contamination fractions, instead, range from below 5\% 
($\sigma_{\varDelta z/(1+z)} = 0.003, 0.01$) to between 10\% and 20\% ($\sigma_{\varDelta z/(1+z)} = 0.03, 0.06$). For low-density environments, $f_{Rec}$ is lower, ranging from around 75\% at all redshifts for $\sigma_{\varDelta z/(1+z)} = 0.003$ and progressively decreasing to slightly above 35\% with increasing photometric redshift uncertainty. Contamination 
fractions are also lower, but show the same trend with $\sigma_{\varDelta z/(1+z)}$ as for high-density environments, being below 5\% for $\sigma_{\varDelta z/(1+z)} = 0.003, 0.01$ and progressively reaching 10-15\% for $\sigma_{\varDelta z/(1+z)} = 0.03, 0.06$.

\subsection{The impact of $n$ in the case of varying $\sigma_{\varDelta z/(1+z)}$}
As done before for the best-case photometric redshift uncertainty of $\sigma_{\varDelta z/(1+z)} = 0.01$, we also investigated the dependence of the accuracy in reconstructing the environment on the parameters of the volume for the various photometric redshift errors. We found that the $R$ and $n$ parameters have a great impact on the measurement of the density field also in the case of large $\sigma_{\varDelta z/(1+z)}$ values. 

\begin{figure}
\resizebox{\hsize}{!}{\includegraphics{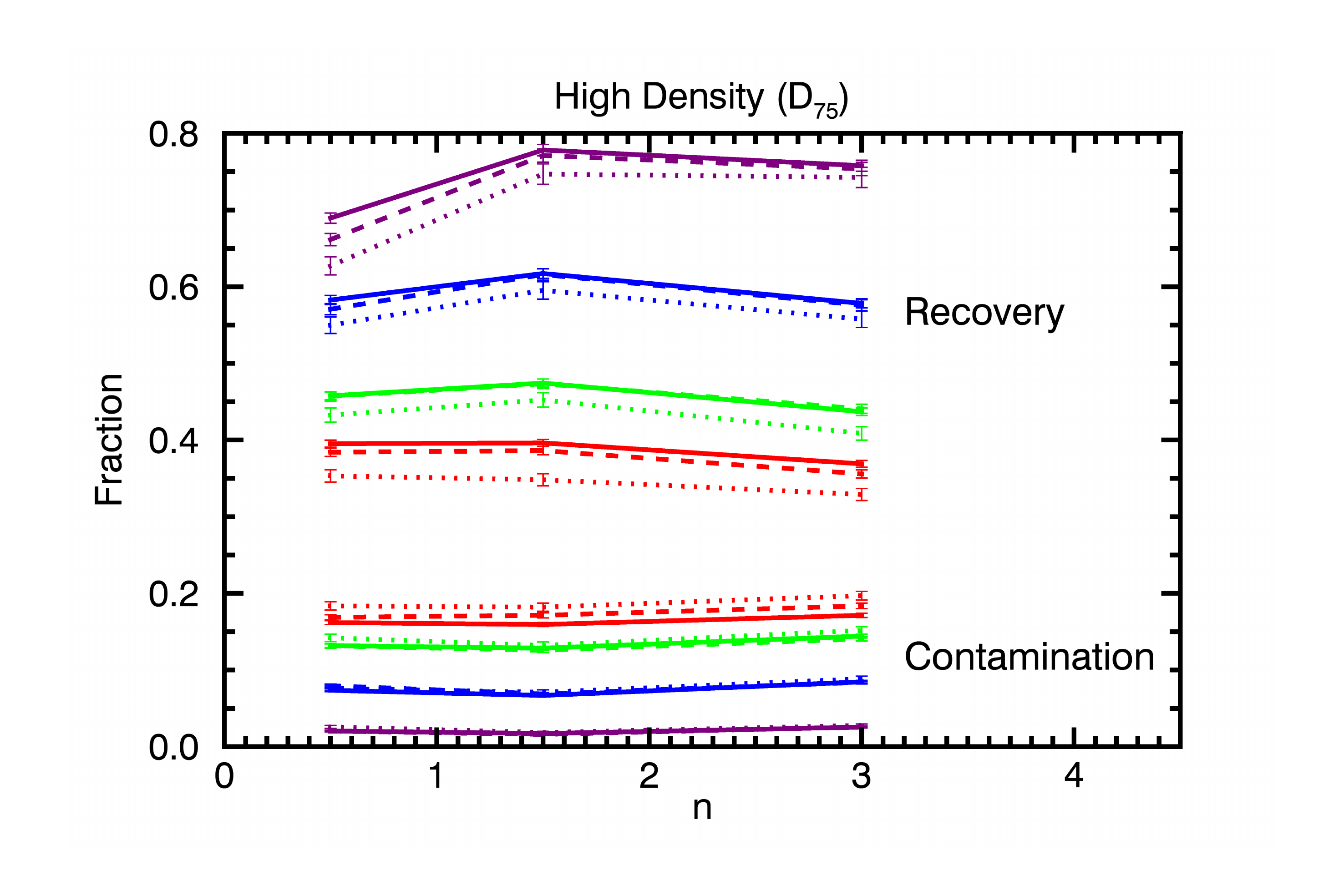}}
\resizebox{\hsize}{!}{\includegraphics{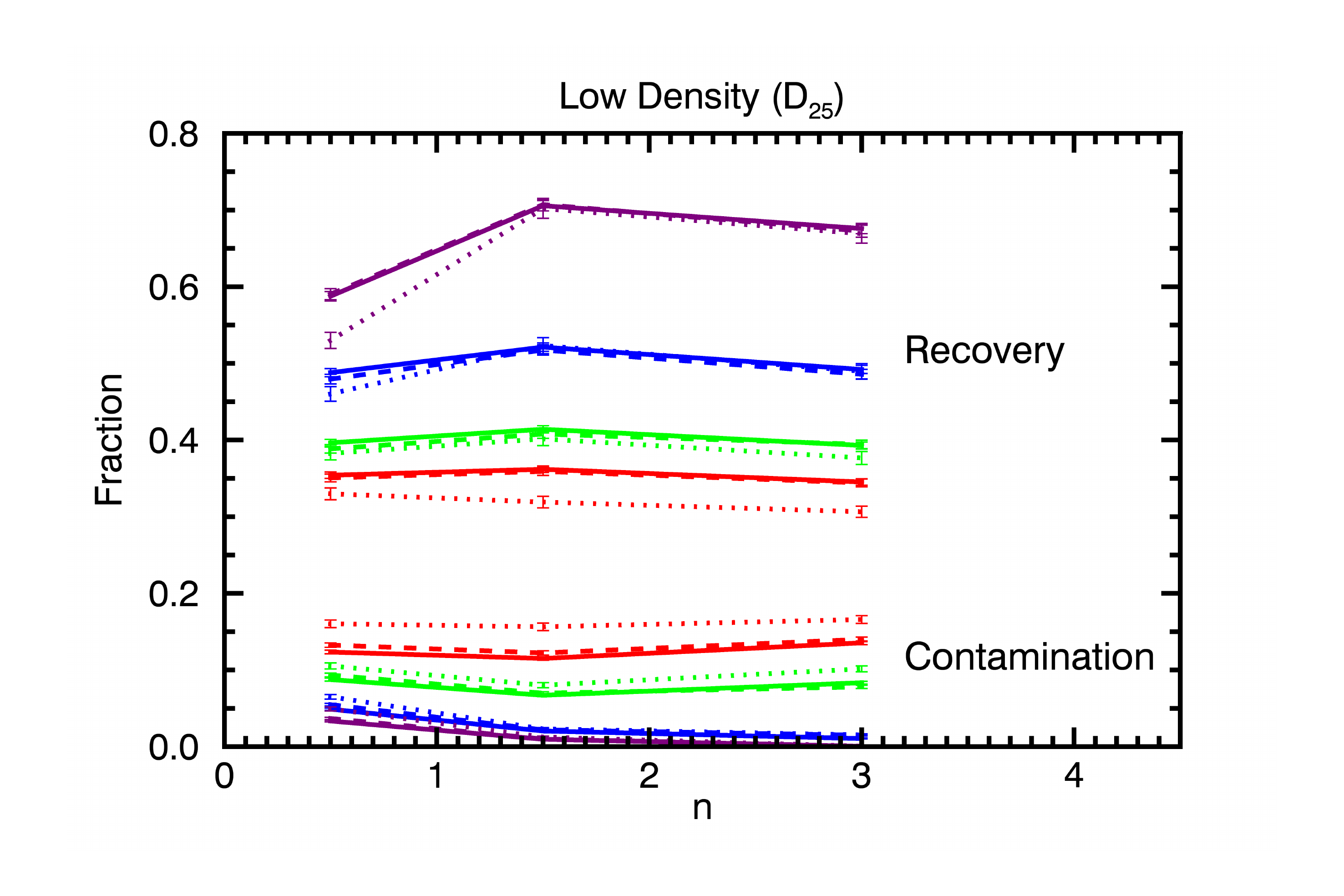}}
\caption{\textit{Varying $\sigma_{\varDelta z/(1+z)}$ - effect of $n$}. This figure shows $f_{Rec}$ and $f_{Con}$ as a function of the $n$ parameter in three redshift bins, namely $1.50 < z < 1.75$ (solid lines), $2.00 < z < 2.25$ (dashed lines) and $2.50 < z < 2.75$ (dotted lines). Bottom panel refers to low-density environments while top panel refers to high-density environments. The various curves are color-coded according to the various values of $\sigma_{\varDelta z/(1+z)}$ (purple: $\sigma_{\varDelta z/(1+z)} = 0.003$, blue: $\sigma_{\varDelta z/(1+z)} = 0.01$, green: $\sigma_{\varDelta z/(1+z)} = 0.03$, and red: $\sigma_{\varDelta z/(1+z)} = 0.06$). The aperture radius has been kept fixed to $R_{R} = R_{T} = 1$ Mpc.}
\label{allsurveyN}
\end{figure}

Figure \ref{allsurveyN} shows $f_{Rec}$ and $f_{Con}$ for only three redshift bins (namely $1.50 \le z \le 1.75$, $2.00 \le z \le 2.25$ and $2.50 \le z \le 2.75$) as a function of the $n$ parameter for various values of $\sigma_{\varDelta z/(1+z)}$. It can be seen how, in the high-density case, $f_{Rec}$ is generally higher for a value of $n = 1.5$, a trend shared by all redshift bins. Only for really large values of the photometric redshift uncertainty ($\sigma_{\varDelta z/(1+z)} = 0.06$) is $n = 0.5$ a viable solution too, but the difference in $f_{Rec}$ between this value and the one obtained with $n = 1.5$ is negligible. Instead, $f_{Con}$ is always lower in the case of $n = 1.5$ independently of redshift and photometric redshift uncertainty.

In the low-density case, shown in the bottom panel of the same figure, it is shown that $f_{Rec}$ is always higher and $f_{Con}$ is lower when a value of $n = 1.5$ is adopted. This generalizes the result obtained previously in the case of $\sigma_{\varDelta z/(1+z)} = 0.01$ (see Figure \ref{fractionssigma}) showing that environmental reconstruction is better performed when it is chosen a value of the cylinder length of a similar size as the $\pm 1.5\sigma$ error on the photometric redshifts.

\subsection{The effect of the fixed aperture radius $R$}
\begin{figure}
\resizebox{\hsize}{!}{\includegraphics{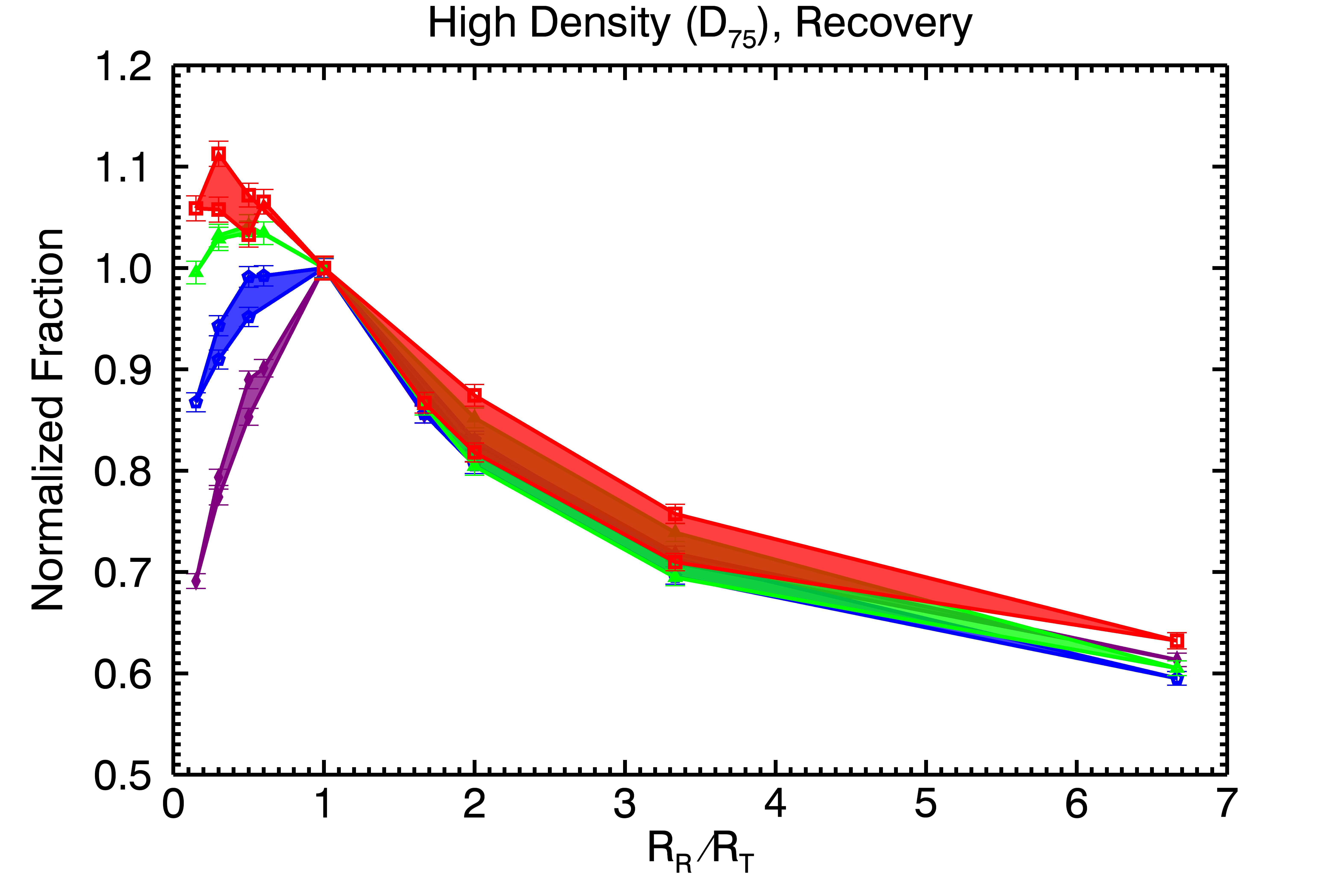}}
\resizebox{\hsize}{!}{\includegraphics{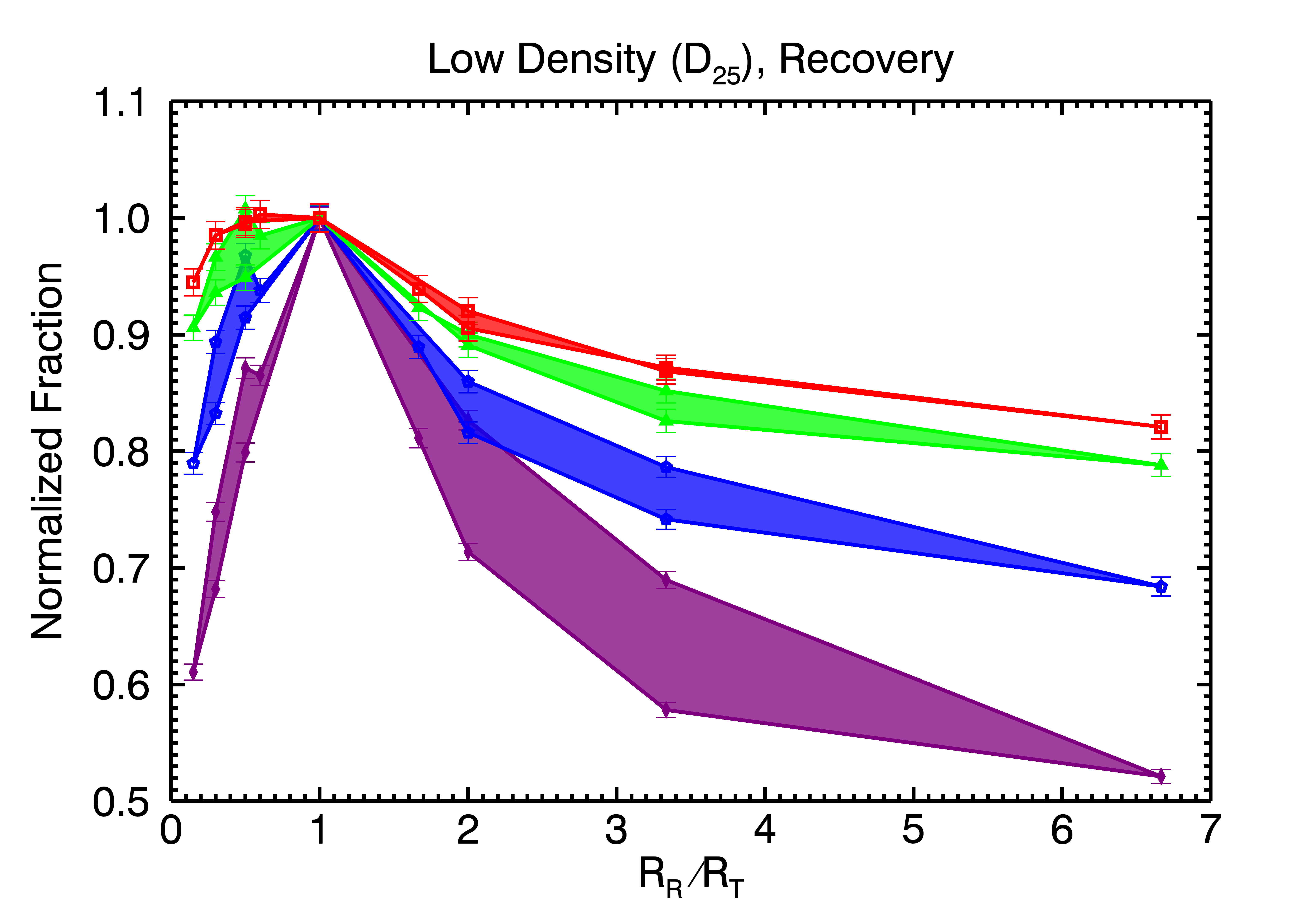}}
\caption{\textit{Varying $\sigma_{\varDelta z/(1+z)}$ - effect of $R_{R}/R_{T}$, Recovery}. This figure shows $f_{Rec}$ (normalized to the value of $f_{Rec}$ at $R_{R}/R_{T} = 1$ separately for each value of $\sigma_{\varDelta z/(1+z)}$) as a function of $R_{R}/R_{T}$ in the redshift bin $1.50 < z < 1.75$. Bottom panel refers to low-density environments while top panel refers to high-density environments. The various curves are color-coded according to the various values of $\sigma_{\varDelta z/(1+z)}$ (purple: $\sigma_{\varDelta z/(1+z)} = 0.003$, blue: $\sigma_{\varDelta z/(1+z)} = 0.01$, green: $\sigma_{\varDelta z/(1+z)} = 0.03$, and red: $\sigma_{\varDelta z/(1+z)} = 0.06$). Shaded regions show the dispersion in the fraction values at fixed $R_{R}/R_{T}$ given by different $R_{T}$, due to the fact that smaller $R_{T}$ are reconstructed better than larger $R_{T}$. The length of the volume has been kept fixed, with $n = 1.5$.}
\label{allsurveyRfrec}
\end{figure}

\begin{figure}
\resizebox{\hsize}{!}{\includegraphics{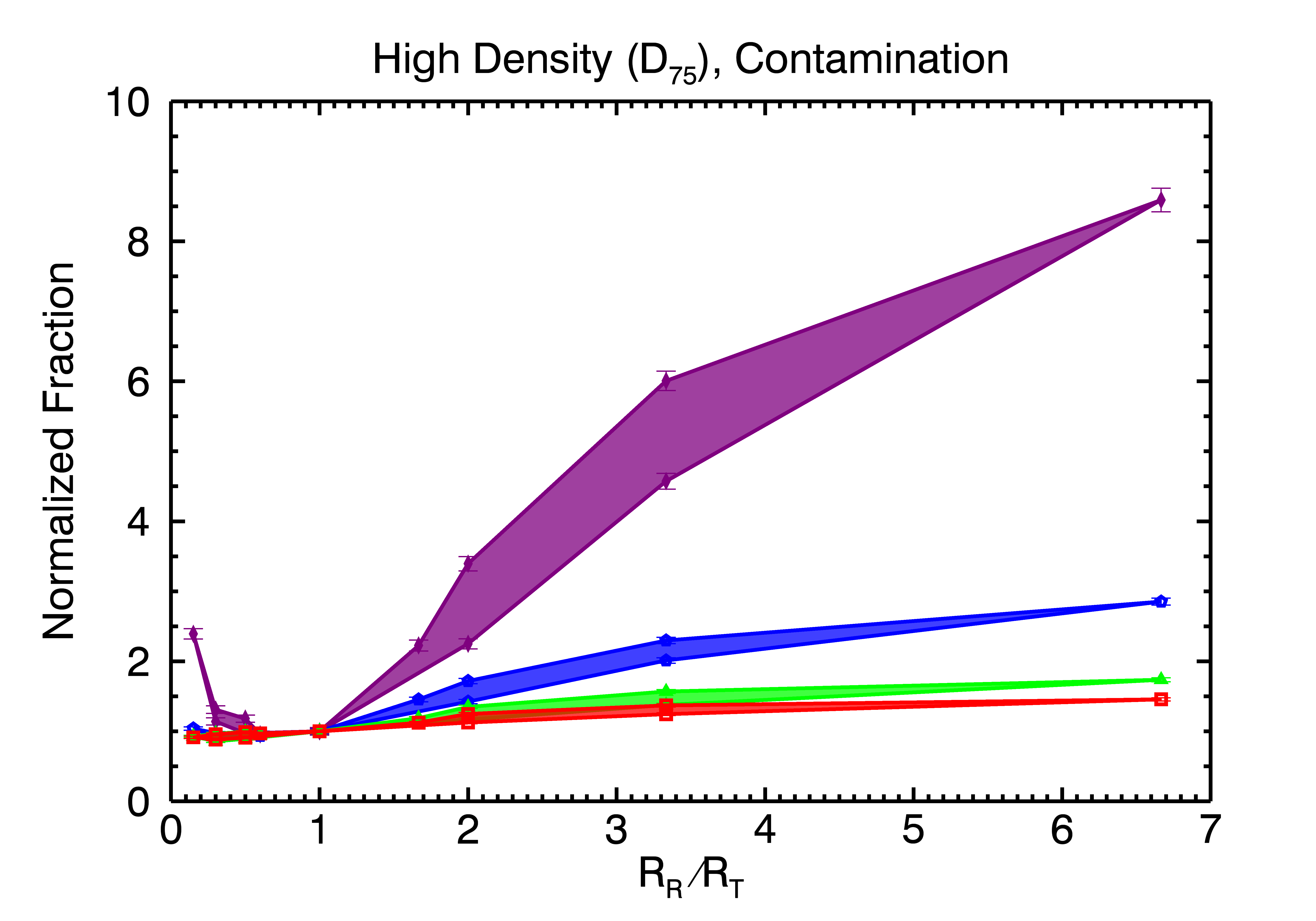}}
\resizebox{\hsize}{!}{\includegraphics{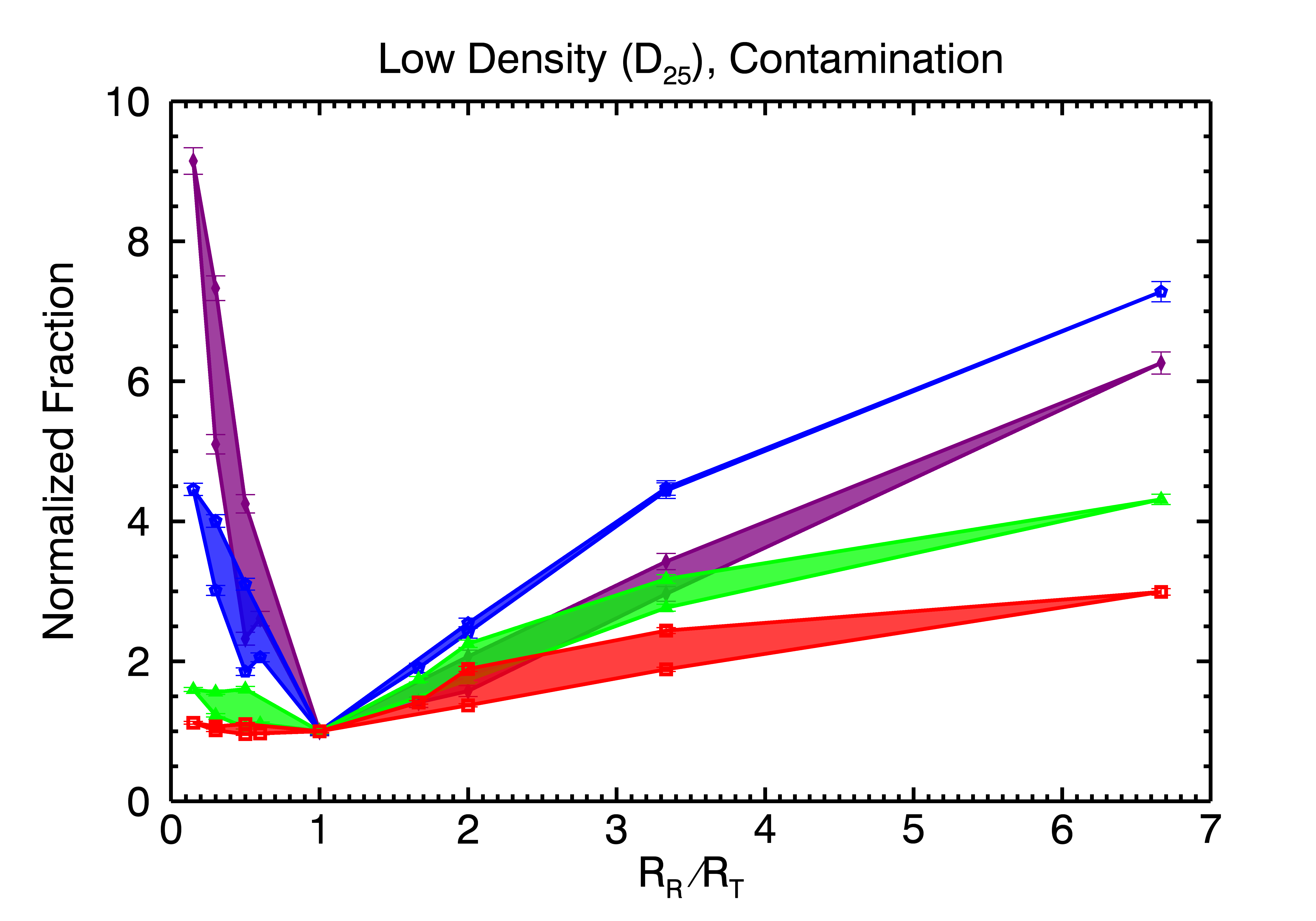}}
\caption{\textit{Varying $\sigma_{\varDelta z/(1+z)}$ - effect of $R_{R}/R_{T}$, Contamination}. This figure shows $f_{Con}$ (normalized to the value of $f_{Con}$ at $R_{R}/R_{T} = 1$ separately for each value of $\sigma_{\varDelta z/(1+z)}$) as a function of $R_{R}/R_{T}$ in the redshift bin $1.50 < z < 1.75$. Bottom panel refers to low-density environments while top panel refers to high-density environments. The various curves are color-coded according to the various values of $\sigma_{\varDelta z/(1+z)}$ (purple: $\sigma_{\varDelta z/(1+z)} = 0.003$, blue: $\sigma_{\varDelta z/(1+z)} = 0.01$, green: $\sigma_{\varDelta z/(1+z)} = 0.03$, and red: $\sigma_{\varDelta z/(1+z)} = 0.06$). Shaded regions show the dispersion in the fraction values at fixed $R_{R}/R_{T}$ given by different $R_{T}$, due to the fact that smaller $R_{T}$ are reconstructed better than larger $R_{T}$. The length of the volume has been kept fixed, with $n = 1.5$.}
\label{allsurveyRfcon}
\end{figure}

Figures \ref{allsurveyRfrec} and \ref{allsurveyRfcon} show $f_{Rec}$ and $f_{Con}$ for only one redshift bin ($1.50 \le z \le 1.75$) as a function of the ratio $R_{R}/R_{T}$, for various values of $\sigma_{\varDelta z/(1+z)}$. In order to better visualize the trend in the fractions with $R_{R}/R_{T}$, we normalized all values of $f_{Rec}$ and $f_{Con}$ to their value at $R_{R}/R_{T} = 1$, separately for every case of $\sigma_{\varDelta z/(1+z)}$ considered. In this way the intrinsic dispersion in the data, due to the fact that smaller scales are better reconstructed than large ones has been reduced for the sake of clarity. We show residual scatter as shaded regions for each curve. The points at $R_{R}/R_{T} < 1$ are given by apertures whose ratio is lower than 1, for example $R_{R} = 1$ Mpc and $R_{T} = 2$ Mpc which yield $R_{R}/R_{T} = 0.5$ and so on. Moreover a fixed value of $R_{R}/R_{T}$ could be given by more than one combination of $R_{R}$ and $R_{T}$, for example $R_{R}/R_{T} = 0.5$ could be given by $R_{R} = 0.3$ Mpc and $R_{T} = 0.6$ Mpc or by $R_{R} = 1$ Mpc and $R_{T} = 2$ Mpc. For this reason at a given value of $R_{R}/R_{T}$ more than one point may be visible. The only exception is the point at $R_{R}/R_{T} = 1$ where all the curves have been normalized to unity.

In this way, the normalization of each curve corresponding to each value of $\sigma_{\varDelta z/(1+z)}$ is lost, but the shapes and the trends with $R_{R}/R_{T}$ can be better studied.

Recovery fractions (Figure \ref{allsurveyRfrec}) clearly show a steep decrease for $R_{R}/R_{T} > 1$ for every value of $\sigma_{\varDelta z/(1+z)}$. This decrease is similar for all values of photometric redshift uncertainty in the $D_{75}$ case and it is shallower for larger values of $\sigma_{\varDelta z/(1+z)}$ in the $D_{25}$ case. For values of $R_{R}/R_{T} < 1$, instead, an increase in the fraction values is present for values of $\sigma_{\varDelta z/(1+z)} < 0.06$ in the $D_{25}$ case, and for values of $\sigma_{\varDelta z/(1+z)} \le 0.01$ in the $D_{75}$ case. For these values of $\sigma_{\varDelta z/(1+z)}$ a maximum in the Recovery fraction can be clearly individuated at $R_{R}/R_{T} = 1$, which translates in the best value for the fixed aperture radius to obtain an accurate environmental reconstruction. For larger $\sigma_{\varDelta z/(1+z)}$ values, instead, smaller apertures than the physical scale which is to be investigated should be considered.

Contamination fractions (Figure \ref{allsurveyRfcon}) show a minimum in $R_{R}/R_{T} = 1$ for all values of $\sigma_{\varDelta z/(1+z)}$ only in the $D_{25}$ case, while the minimum is clearly recognizable in the $D_{75}$ case only for values of $\sigma_{\varDelta z/(1+z)} \le 0.01$. For larger values of the photometric redshift uncertainty the curves corresponding to the Contamination fractions in the high-density case are rather flat or slightly increasing. Again, for small values of $\sigma_{\varDelta z/(1+z)}$ the trend is confirmed of $R_{R}/R_{T} = 1$ being the best choice for the environmental reconstruction, while for larger values, an aperture radius smaller than the physiscal scale that is to be investigated is probably preferable. This generalizes the result obtained previously in the case of $\sigma_{\varDelta z/(1+z)} = 0.01$ (see Figure \ref{fractionsradius}).

The fact that the best environmental reconstruction is obtained for $R_{R} \simeq R_{T}$ is not an obvious result, as for large photometric redshift uncertainties fixed aperture radii smaller than the physical scale of the studied environment may be the best option to limit the number of contaminating interlopers. Here we showed that this is indeed the case for $\sigma_{\varDelta z/(1+z)} = 0.03, 0.06$, while for lower values of $\sigma_{\varDelta z/(1+z)}$ the situation where $R_{R}/R_{T} = 1$ is the one that optimizes the environmental reconstruction. For all curves, shaded regions show the residual dispersion (after normalization) in the fraction values at fixed $R_{R}/R_{T}$. This dispersion is due to the fact that smaller scales ($R_{R} = R_{T} = 0.3$ Mpc) are reconstructed better than the large ones ($R_{R} = R_{T} = 2$ Mpc). Therefore, even at fixed $R_{R}/R_{T}$ smaller scales will have higher $f_{Rec}$ and lower $f_{Con}$ compared to larger scales, as already found for the $\sigma_{\varDelta z/(1+z)} = 0.01$ case.

To summarize, we can conclude that it is possible to reconstruct environment in an accurate way only if the photometric redshift uncertainty is small ($\sigma_{\varDelta z/(1+z)} \le 0.01$), otherwise the environment will not be sufficiently recovered ($f_{Rec} < 50\%\div 60\%$) and it will become too contamined ($f_{Con} > 10\%$). Moreover, for uncertainty values $\sigma_{\varDelta z/(1+z)} \le 0.01$ the best environmental reconstruction will be obtained for $n = 1.5$ and for $R_{R}/R_{T} = 1$. This remains generally true also for values of $\sigma_{\varDelta z/(1+z)} = 0.03$, although the Recovery fraction is lower and the Contamination fraction is higher. For values of $\sigma_{\varDelta z/(1+z)} > 0.03$, the Recovery and Contamination fractions do not allow to reconstruct environment in an accurate fashion and volume parameters $n = 0.5$ and $R_{R}/R_{T} < 1$ are the ones that optimize the measurement of the density field as they limit the number of contaminating interlopers.

\section{The reconstruction of the Galaxy Stellar Mass Function for the best-case $\sigma_{\varDelta z/(1+z)} = 0.01$}
\label{mfrec}
On the basis of this analysis we now investigate whether the accuracy of the environmental reconstruction with photometric redshifts has consequences on the differential study of galaxy stellar mass functions in different environments. It is in fact known, from spectroscopic surveys, that mass functions of galaxies in different environments have a different shape at least up to $z \le 1$ (\citeads{2010A&A...524A..76B}). This is caused by the different formation and evolution scenarios of galaxies within clusters and in the field. For this reason, the contamination of interlopers from different environments (the $f_{Con}$ defined above), together with the dilution of the signal of environmental segregation (due to values of $f_{Rec} < 100\%$) might have the effect of changing the shape of the mass function, by mixing together galaxies with different properties. In the following we attempt an investigation to quantify the degree of accuracy with which an analysis of the GSMF in different environments is possible if the density field is reconstructed using photometric redshifts with an error of $\sigma_{\varDelta z/(1+z)} = 0.01$. We chose this value of photometric redshift uncertainty as for higher values it is not possible to reliably reconstruct galaxy environments as described in section \ref{photozerror}. Moreover, this value is the one that has been derived for the photometric redshifts of the UltraVISTA survey, to which we plan to apply the results of this work. We chose this value as it is in agreement with the one reported in Figure 1 of \citetads{2013A&A...556A..55I}, which shows a comparison between the photometric redshifts and a sample of spectroscopic redshifts at $K_S \le 24$. It is also in agreement with the mean of the error values reported in Table 1 of \citetads{2013A&A...556A..55I}, weighted by the number of sources in each spectroscopic sample used to determine the error. This takes into account the fact that the spectroscopic samples used to derive the values of Table 1 of \citetads{2013A&A...556A..55I} are sometimes small, composed of a few tens of galaxies, and therefore the errors reported may not be representative of the whole spectroscopic sample at $K_S \le 24$.

\subsection{The lightcone mock catalogue GSMF}
\label{mfmocks}
\begin{figure*}
\centering
\includegraphics[scale=0.75]{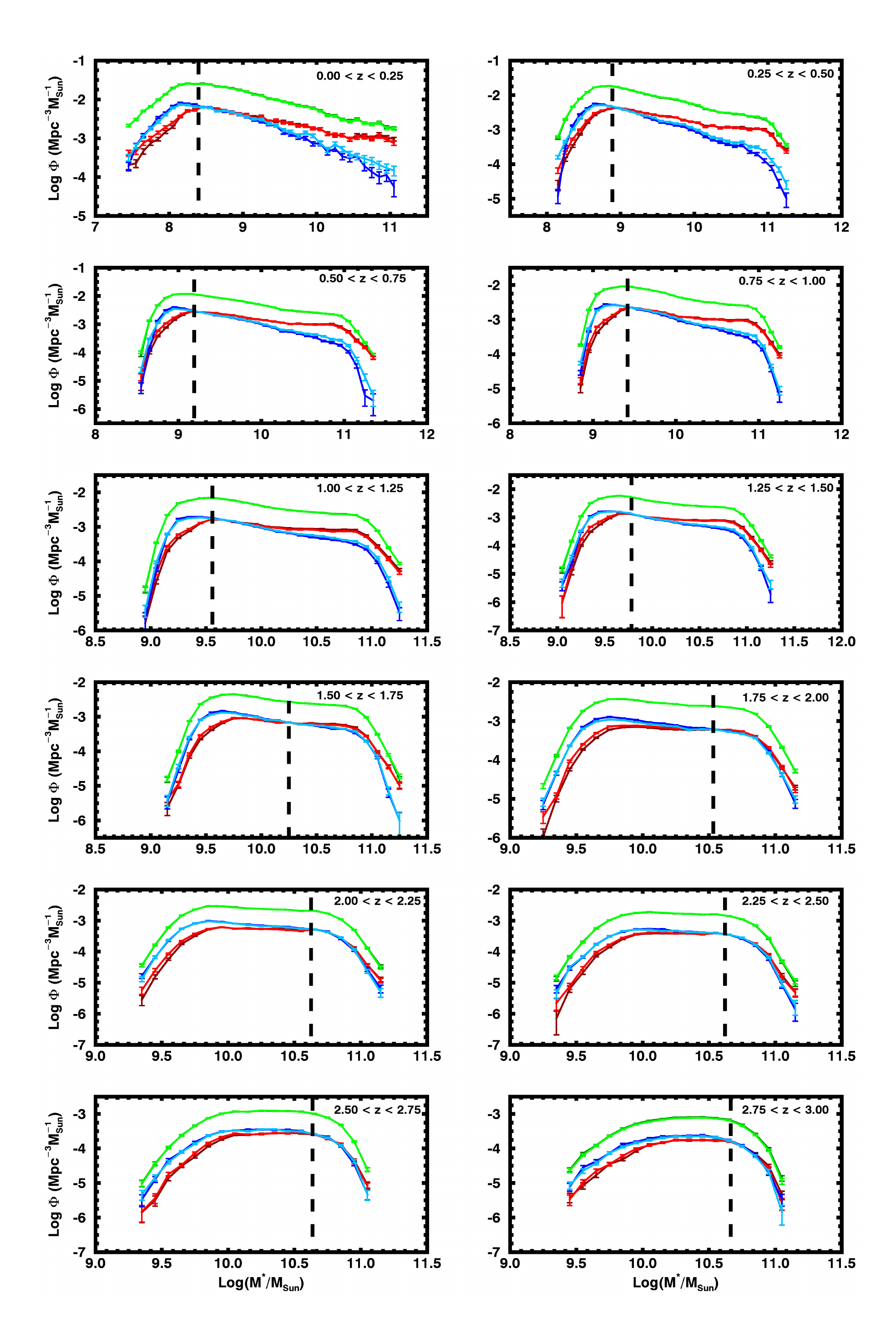}
\caption{\textit{Mock Catalogues Mass Functions - $R_{R} = R_{T} = 0.3$ Mpc}. The dark green and light green curves refer to the total GSMF (respectively using $z_{true}$ and $z_{phot}$ for the calculation of the mass function). Dark red and light red curves refer to high-density environments (respectively using $\varrho_{true}$, dark red, and using $\varrho_{rec}$, light red). Dark blue and light blue curves refer to low-density environments (respectively using $\varrho_{true}$, dark blue, and using $\varrho_{rec}$, light blue). The black dashed lines are the mass completeness limits described in the text. The parameter values for the aperture are set to $R_{R} = R_{T} = 0.3$ and $n = 1.5$, with $\sigma_{\varDelta z/(1+z)} = 0.01$. Error bars refer to $1/V_{max}^2$ estimates.}
\label{mfmocks03}
\end{figure*}

In this section we compare the mass functions for the galaxies of the mock catalogue used for the environmental analysis. In particular, we derived the mass functions of high-density and low-density environments based on $\varrho_{rec}$ with $n = 1.5$, $\sigma_{\varDelta z/(1+z)} = 0.01$ and $R_{T} = R_{R} = 0.3$ and $R_{T} = R_{R} = 2$ Mpc. Only the smallest and the largest radii have been considered since they are those who grant the best and the worst environmental reconstruction. All other values of $R$ will grant intermediate $f_{Rec}$ and $f_{Con}$. The mass functions of the mock catalogues are shown in Figure \ref{mfmocks03} for the $R_{R} = R_{T} = 0.3$ Mpc case and in Figure \ref{mfmocks2} for the $R_{R} = R_{T} = 2$ Mpc case.

We derived mass functions using the non-parametric $1/V_{max}$ estimator (see \citeads{1980ApJ...235..694A}, \citeads{2002A&A...395..443B} and references therein for further information), considering all the galaxies down to $K \le 24$. As the $z_{max}$ information was not available, we used the total volume in the considered redshift range. In particular, not taking the $z_{max}$ into account will affect maily the low-mass end of the mass function, where the volume occupied by each source is more likely to be overestimated. For this reason, we also derived mass completeness limits as in \citetads{2010A&A...523A..13P} but at the upper boundary of each redshift bin (instead than at the lower one, which is the case when $z_{max}$ values are available).

The error bars shown in the plots represent only the Poissonian errors. As the UltraVISTA Survey field is smaller compared to the area of the mock catalogues that we considered ($1.48\: \deg^2$ for UltraVISTA compared to the $8\: \deg^2$ used in this work), we have calculated the GSMF also for galaxies in five independent areas of $1.48\: \deg^2$ to simulate the real UltraVISTA data. We extracted five independent areas and have redone the GSMF calculation in each of them in order to not be biased by cosmic variance. We found that, despite the larger error bars due to the lower number of galaxies present in the smaller fields, the results found with the larger area and exposed below hold up to $z \sim 2$. At higher redshifts, massive galaxies of $M^{\ast} \gtrsim 10^{11} M_{\odot}$ (which carry most of the signal of environmental difference) are too few in every redshift bin, due to the smaller area, already in the \textit{True} environment case; therefore no environmental difference is recoverable. Nevertheless this may not be a limit in the analysis of the UltraVISTA data, as different mass or redshift bins may be applied to increase statistics, and because mock galaxy catalogues may underestimate the number of massive galaxies in comparison to reality (see Figures 14 and 15 of \citeads{2013A&A...556A..55I}). For these reasons we show here GSMF calculated with an area of $8\: \deg^2$.

It is known that galaxies in high-density environments occupy a smaller volume than galaxies in low-density environments. To account for this fact and be able to compare GSMF in different environments, we normalized GSMF in high-density 
and low-density environments to $1/4$ of the value that the total GSMF has at the mass limit, both in the \textit{True} and \textit{Reconstructed} environments. In this way, although the information on the GSMF normalization is lost, it is still possible to compare their shapes.

\begin{figure*}
\centering
\includegraphics[scale=0.775]{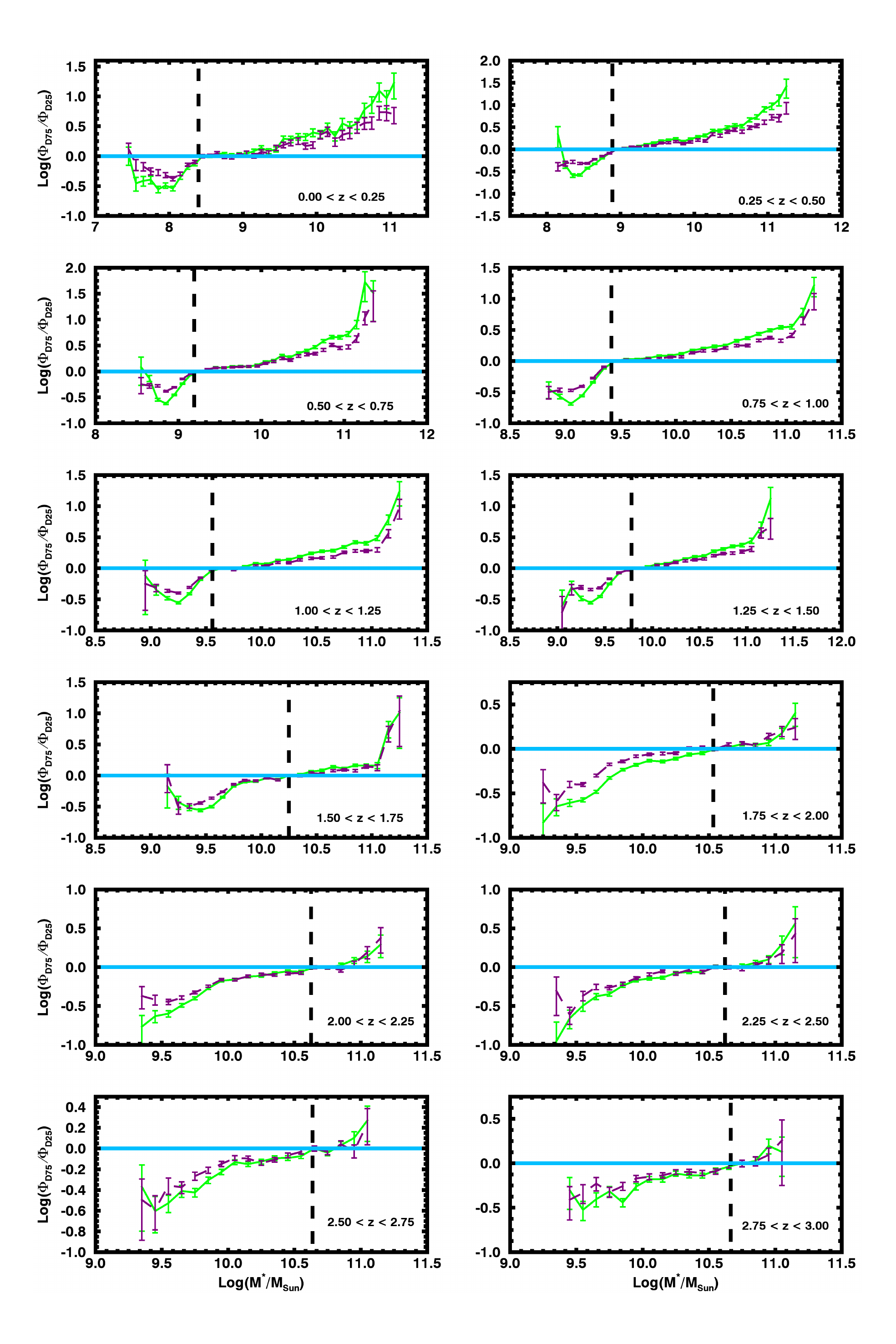}
\caption{\textit{Ratio of high-density and low-density mass functions - $R_{R} = R_{T} = 0.3$ Mpc}. Ratio of the high-density mass function and the low-density one ($\Phi_{D75}/\Phi_{D25}$) in the \textit{True} (light green curve) and \textit{Reconstructed} (purple curve) environments. The black dashed lines are the mass completeness limits described in the text. The parameter values for the aperture are set to $R_{R} = R_{T} = 0.3$ Mpc and $n = 1.5$, with $\sigma_{\varDelta z/(1+z)} = 0.01$. Error bars refer to $1/V_{max}^2$ estimates.}
\label{mfmocksratio03}
\end{figure*}

\begin{figure*}
\centering
\includegraphics[scale=0.75]{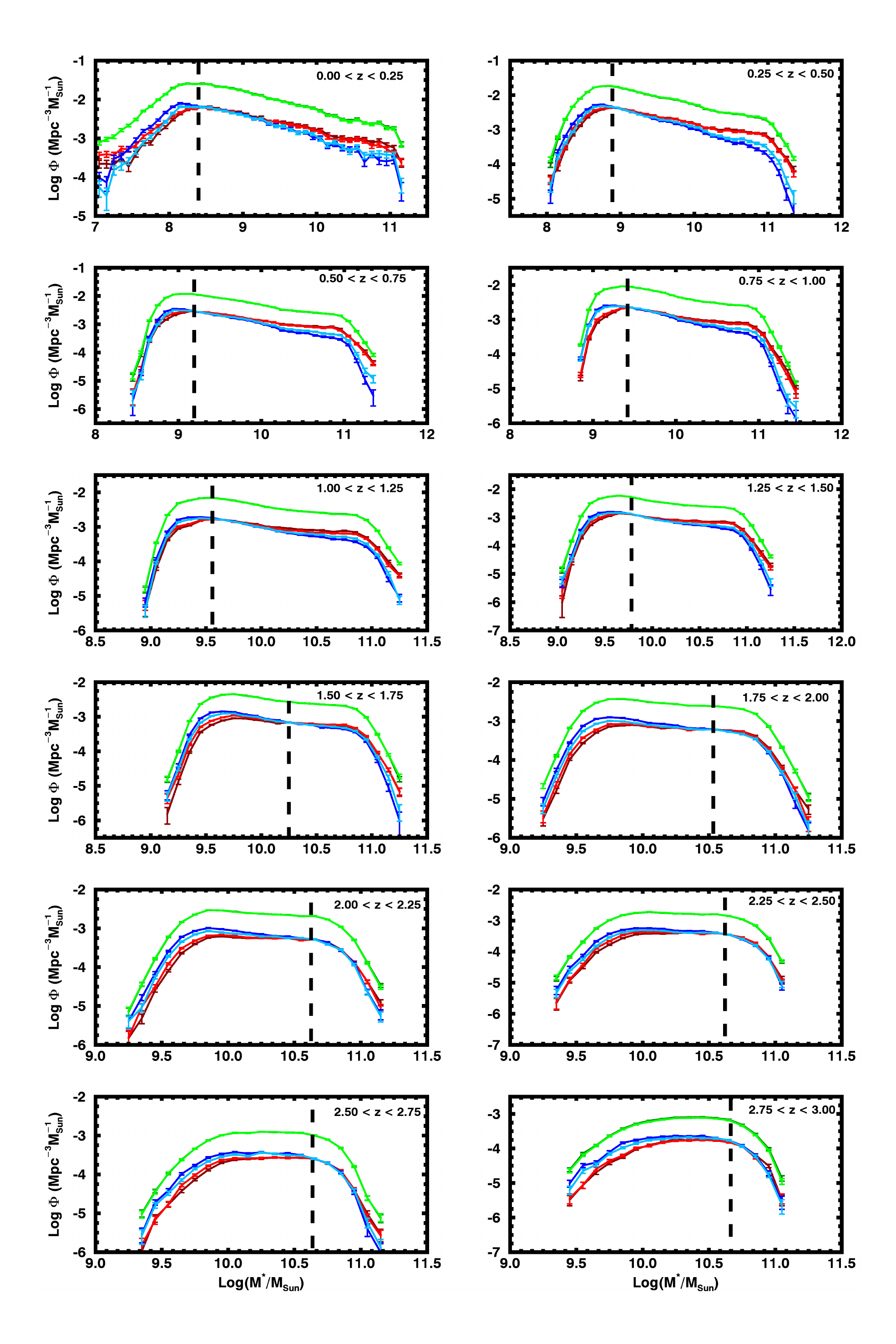}
\caption{\textit{Mock Catalogues Mass Functions - $R_{R} = R_{T} = 2$ Mpc}. The dark green and light green curves refer to the total GSMF (respectively using $z_{true}$ and $z_{phot}$ for the calculation of the mass function). Dark red and light red curves refer to high-density environments (respectively using $\varrho_{true}$, dark red, and using $\varrho_{rec}$, light red). Dark blue and light blue curves refer to low-density environments (respectively using $\varrho_{true}$, dark blue, and using $\varrho_{rec}$, light blue). The black dashed lines are the mass completeness limits described in the text. The parameter values for the aperture are set to $R_{R} = R_{T} = 2$ Mpc and $n = 1.5$, with $\sigma_{\varDelta z/(1+z)} = 0.01$. Error bars refer to $1/V_{max}^2$ estimates.}
\label{mfmocks2}
\end{figure*}

From Figure \ref{mfmocks03} it is possible to see how the mass functions of \textit{True} $D_{75}$ and $D_{25}$ environments are intrinsically different in the $R_{R} = R_{T} = 0.3$ Mpc case. This difference is a function of mass and redshift and for masses $M \gtrsim 10^{11} M_{\sun}$ ranges between $\gtrsim 1$ dex at $z < 0.5$ and $\sim 0.3$ dex at $z \sim 2.5$. At higher redshifts the two mass functions become undistinguishable. This trend with mass and redshift of the differences between \textit{True} environment GSMF can be better appreciated in the ratio between $D_{75}$ and $D_{25}$ GSMF, shown in Figure \ref{mfmocksratio03}.

\begin{figure*}
\centering
\includegraphics[scale=0.775]{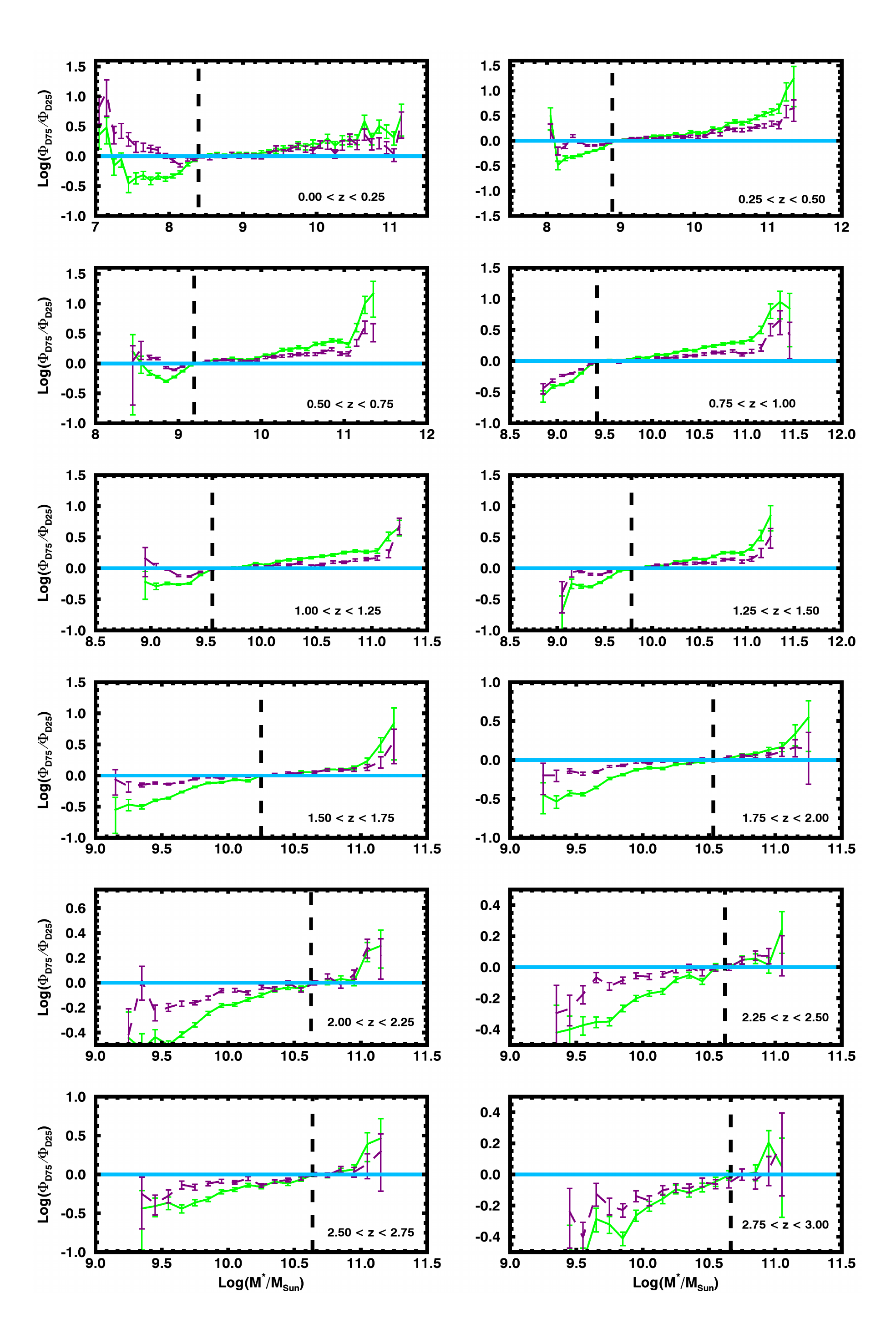}
\caption{\textit{Ratio of high-density and low-density mass functions - $R_{R} = R_{T} = 2$ Mpc}. Ratio of the high-density mass function and the low-density one ($\Phi_{D75}/\Phi_{D25}$) in the \textit{True} (light green curve) and \textit{Reconstructed} (purple curve) environments. The black dashed lines are the mass completeness limits described in the text. The parameter values for the aperture are set to $R_{R} = R_{T} = 2$ Mpc and $n = 1.5$, with $\sigma_{\varDelta z/(1+z)} = 0.01$. Error bars refer to $1/V_{max}^2$ estimates.}
\label{mfmocksratio2}
\end{figure*}

The situation is worse when considering $R_{R} = R_{T} = 2$ Mpc (Figures \ref{mfmocks2} and \ref{mfmocksratio2}). Already at low redshifts the difference between the \textit{True} $D_{75}$ and $D_{25}$ GSMF is $\lesssim 0.5$ dex for masses below $10^{11} M_{\sun}$. This difference is of $\sim 0.1$ dex at $z \sim 2.5$ for $M \sim 10^{11} M_{\sun}$. The ratio of the two mass functions is significantly different from one below $z \sim 2$ and only for masses above $10^{11} M_{\sun}$.

In the same four figures we also report GSMF in different \textit{Reconstructed} environments, together with their ratios. In the $R_{R} = R_{T} = 0.3$ Mpc case (Figure \ref{mfmocks03}) it is possible to see how the \textit{Reconstructed} environment GSMF follow closely the \textit{True} ones, the $D_{75}$ being well distinguishable from the $D_{25}$ at low redshifts and showing the same trends with mass and redshift. Also the ratio of the $D_{75}$ to the $D_{25}$ GSMF (Figure \ref{mfmocksratio03}) follows the \textit{True} case, although it is possible to see that the differences result somewhat damped when passing from $\varrho_{true}$ to $\varrho_{rec}$. To quantify this decrease of the difference between high-density and low-density environments we report in Table \ref{maxdiff03} the maximum decrease of the ratio of the mass functions in extreme environments between the $\varrho_{true}$ and the $\varrho_{rec}$ case, defined as
\begin{equation}\label{xi}
\xi = \max \left[\log\left(\frac{\Phi(D75)}{\Phi(D25)}\right)\Bigg|_{True} - \log\left(\frac{\Phi(D75)}{\Phi(D25)}\right)\Bigg|_{Rec} \right]  
\end{equation}

\begin{table}
\caption{Maximum decrease between $\varrho_{rec}$ and $\varrho_{true}$ $D_{75}$ and $D_{25}$ GSMF. The ratio $\xi$ is calculated as in Equation \eqref{xi}, $M_{\xi}$ is the mass at which $\xi$ is located. These values refer to $R_{R} = R_{T} = 0.3$ Mpc.}
\label{maxdiff03}
\centering
\begin{tabular}{c c c}
\hline\hline
Redshift            & $\xi$ (dex) & $\log(M_{\xi}/M_{\sun})$ \\
\hline
$0.00 < z < 0.25$   & 0.53        & 11.05       \\
$0.25 < z < 0.50$   & 0.48        & 11.25       \\
$0.50 < z < 0.75$   & 0.68        & 11.25       \\
$0.75 < z < 1.00$   & 0.24        & 11.25       \\
$1.00 < z < 1.25$   & 0.26        & 11.25       \\
$1.25 < z < 1.50$   & 0.45        & 11.25       \\
$1.50 < z < 1.75$   & 0.09        & 10.85       \\
$1.75 < z < 2.00$   & 0.16        & 11.15       \\
$2.00 < z < 2.25$   & 0.06        & 10.85       \\
$2.25 < z < 2.50$   & 0.14        & 11.15       \\
$2.50 < z < 2.75$   & 0.13        & 10.95       \\
$2.75 < z < 3.00$   & 0.10        & 10.95       \\
\hline
Average             & 0.28             \\ 
\hline
\end{tabular}
\end{table}

It can be seen how the reduction in the differences is always below $\sim 0.7$ dex. This value is obtained at high masses ($M \sim 10^{11} M_{\sun}$) where the starting (\textit{True} environment) GSMF are intrinsically different. Table \ref{maxdiff2} shows that this is true also for the $R_{R} = R_{T} = 2$ Mpc case, although from Figures \ref{mfmocks2} and \ref{mfmocksratio2} it is possible to see that already the starting \textit{True} environment GSMF are less different between high- and low-density than in the $R_{R} = R_{T} = 0.3$ Mpc case.

\begin{table}
\caption{Maximum decrease between $\varrho_{rec}$ and $\varrho_{true}$ $D_{75}$ and $D_{25}$ GSMF. The ratio $\xi$ is calculated as in Equation \eqref{xi}, $M_{\xi}$ is the mass at which $\xi$ is located. These values refer to $R_{R} = R_{T} = 2$ Mpc.}
\label{maxdiff2}
\centering
\begin{tabular}{c c c}
\hline\hline
Redshift            & $\xi$ (dex) & $\log(M_{\xi}/M_{\sun})$ \\
\hline
$0.00 < z < 0.25$   & 0.28        & 10.85       \\
$0.25 < z < 0.50$   & 0.59        & 11.35       \\
$0.50 < z < 0.75$   & 0.63        & 11.35       \\
$0.75 < z < 1.00$   & 0.42        & 11.45       \\
$1.00 < z < 1.25$   & 0.28        & 11.15       \\
$1.25 < z < 1.50$   & 0.34        & 11.25       \\
$1.50 < z < 1.75$   & 0.29        & 11.25       \\
$1.75 < z < 2.00$   & 0.41        & 11.25       \\
$2.00 < z < 2.25$   & 0.08        & 11.15       \\
$2.25 < z < 2.50$   & 0.15        & 11.05       \\
$2.50 < z < 2.75$   & 0.26        & 11.05       \\
$2.75 < z < 3.00$   & 0.17        & 10.95       \\
\hline
Average             & 0.33 \\
\hline
\end{tabular}
\end{table}

Thus, we showed that if differences are present they will be recovered (although somewhat reduced) while if there are no differences in the \textit{True} environment GSMF then no spurious ones will be introduced when using \textit{Reconstructed} environments. We can then conclude that an analysis of the GSMF in different environments is possible, even when relying on photometric redhsifts for the environmental reconstruction. This result is encouraging, as when using real data the risk will be of missing or underestimating differences in the GSMF of different environments, rather than detecting differences that are not real. This is generally true at all redshifts (up to $z \sim 2.5$), at all masses, and for both large and small scales.

These results allow to draw important conclusions on what to expect from real data. In particular, when investigating the differences between GSMF calculated for galaxies in high-density and low-density environments using high-precision 
photometric redshifts, all differences found may be considered as lower limits of the true differences in galaxy properties. In fact, our analysis shows that the effect of the error of photometric redshifts on the GSMFs of galaxies in different environments is to damp differences between high-density and low-density regions, while nevertheless not deleting them completely. Any environmental effect recovered would be greater if a measure of the true density field were available.

\section{Conclusions and summary}
\label{conclusions}
The GSMF and its relation to different environments are of vital importance for understanding how galaxies have formed and evolved. Galaxy stellar mass functions in different environments have been studied at $z \le 1$ through the use of spectroscopic redshifts and need now to be investigated at higher redshifts. As spectroscopic redshifts are not available for large samples of galaxies, to high redshifts, and on wide sky areas, it is often necessary to rely on photometric redshifts. Photometric redshifts are more easily available for large statistical samples on wide sky areas and in a large resdshift range, but are characterized by a high uncertainty, which may limit their use for deriving GSMF in different environments.

In this work we made use of the validated mock galaxy catalogues by \citetads{2013MNRAS.429..556M} to investigate how the galaxy environment can be reconstructed based on high-precision photometric redshifts (with $\sigma_{\varDelta z/(1+z)} = 0.01$). We selected the mock glaxy sample to have $K \le 24$ and we extracted an area of $8 \deg^2$ from the original $100 \deg^2$ of the catalogue. We used each galaxy's cosmological redshift ($z_{true}$) and we simulated a set of photometric redshifts ($z_{phot}$) with varying precision by adding a Gaussian error to each galaxy's observed redshift (\textit{i.e.} the cosmological redshift to which the contribution of the galaxy peculiar velocity has been added). We chose an error on the photometric redshifts of $\sigma_{\varDelta z/(1+z)} = 0.01$ as a reference one because it is in agreement with the value reported in Figure 1 of \citetads{2013A&A...556A..55I} (which shows a comparison between photometric and spectroscopic redshifts at $K_S \le 24$) and with the mean of the errors reported in Table 1 of \citetads{2013A&A...556A..55I} weighted by the number of sources in each spectroscopic sample used for the calculation (which, due to the fact that they are rather small, may not be representative of the whole spectroscopic sample at $K_S \le 24$).

We estimated galaxy environments through the use of a fixed aperture method, by counting objects inside a cylinder of base radius $R$ and length $h$ proportional to the photometric redshift uncertainty through the parameter $n$ as $h = \pm n \cdot \sigma_{\varDelta z/(1+z)} \cdot (1+z)$. We defined high-density ($D_{75}$) and low-density ($D_{25}$) environments using the 75th and 25th percentiles of the volume density distribution. For each galaxy we derived both a \textit{True} environment estimate ($\varrho_{true}$, using each galaxy's $z_{true}$) and a \textit{Reconstructed} one ($\varrho_{rec}$ using each galaxy's $z_{phot}$). We studied several combinations of both the fixed aperture volume parameters $n$ and $R$ and of the photometric redshift uncertainty $\sigma_{\varDelta z/(1+z)}$. We then compared the derived $\varrho_{true}$ and $\varrho_{rec}$ to study how the density field can be reconstructed with photometric redshifts. We also applied our results to the study of the GSMF in different environments for the best-case photometric redshift uncertainty $\sigma_{\varDelta z/(1+z)} = 0.01$. What we found can be summarized as follows:
\begin{enumerate}
\item
Only using high-precision photometric redshifts ($\sigma_{\varDelta z/(1+z)} = 0.01$) it is still possible to reconstruct galaxy environment in an accurate way. In particular, in order to well recover high- and low-density environments (with $f_{Rec} \geq 60\% \div 80\%$) with a low level of contaminating interlopers ($f_{Con} \leq 10\%$), it is necessary to carefully tune the parameters of the volume used for the estimate of the density field. In our case, the best environmental reconstruction is obtained considering a cylinder with length $\propto \pm 1.5\sigma$ error on the photometric redshift and a radius $R_{R} = R_{T}$. A volume with a length too large or too small compared to the $\pm 1.5\sigma$ error and with a base area too large or too small compared to the size of the physical scale of the studied environment will lead to an inaccurate environmental reconstruction, with lower $f_{Rec}$ and higher $f_{Con}$.

\item
Even if all the volume parameters are tuned so to have the best case of environmental reconstruction, still Recovery fractions are higher ($f_{Rec} \geq 70\%$) and Contamination fractions are lower $f_{Con} \leq 5\%$ for smaller physical scales ($R_{R} = R_{T} = 0.3 \div 0.6$ Mpc) compared to larger ones ($R_{R} = R_{T} = 1 \div 2$ Mpc).

\item
The above results hold well only for high-precision photometric redshifts with $\sigma_{\varDelta z/(1+z)} \leq 0.01$, where Recovery fractions are between $60\div 80\%$ and Contamination fractions below $\leq 10\%$. For higher uncertainty values ($\sigma_{\varDelta z/(1+z)} \geq 0.03$) Recovery fractions lower rapidly to $f_{Rec} < 50\%$ and Contamination fractions increase up to $f_{Con} \sim 20\%$. This result is reasonable if we consider that the typical velocity dispersion inside the richest galaxy clusters is of the order of $\varDelta z \simeq \pm \frac{1500 km/s}{c} \cdot (1+z) \simeq \pm 0.005\cdot (1+z)$, which is comparable to our best-case photometric redshift uncertainty. Moreover, for photometric redshift errors of the order of $\sigma_{\varDelta z/(1+z)} \le 0.01$ still Recovery fractions are higher and Contamination fractions lower for $n = 1.5$ and $R_{R}/R_{T} = 1$, while for higher values of $\sigma_{\varDelta z/(1+z)}$ values of $R_{R}/R_{T} < 1$ are preferable.

\item 
Using only brightest objects ($K \le 22$) as both targets and tracers of the density field, in order to simulate the case of surveys shallower than the UltraVISTA one, has the effect of reducing the maximum redshift to which our analysis can be extended, from $z \le 2.5$ to $z \le 1.5$. Nevertheless, in this redshift range our method is robust enough to reproduce the same results as in the case of $K \le 24$.

\item
When applying these results to the GSMF (in the best-case of $\sigma_{\varDelta z/(1+z)} = 0.01$) it is found that differences (if present) can be recovered accurately, although some reduction (which reaches at most $\sim 0.7$ dex, with an average of $\sim 0.3$ dex) is inevitable. Nevertheless, fictitious differences do not seem to be created, therefore any environmental segregation found in real data may be regarded as a lower limit of what would be found if a measure of the intrinsic density field were available.
\end{enumerate}

With this study it has been found that an analysis of the GSMF in different environments is possible only with high-precision $\sigma_{\varDelta z/(1+z)} \le 0.01$ photometric redshifts, provided that the fixed aperture lenght and radius are optimized to give the best measurement of the density field. Galaxy stellar mass functions in different environments can be studied while keeping in mind that photometric redshifts, even high-precision ones, reduce differences between high-density and low-density environments by as much as $\sim 0.7$ dex. Further investigation of these issues is strongly needed as the availability of large area, deep photometric redshift surveys is increasing. Photometric redshifts allow the measurement of galaxy properties on large sky areas and in a large redshift range, but it is necessary to carefully check for the effects that their uncertainty has on the analysis that is going to be performed. This work constitutes a preliminary study in order to better understand results found investigating real data. As an application of this method we plan an analysis of the GSMF of the UltraVISTA survey \citepads{2012A&A...544A.156M} in different environments and for this reason all the possible parameters of this study have been tuned so to match those of the sample that will be used \citepads{2013A&A...556A..55I}. The final aim of this work is to try to better understand the systematic effects in the reconstruction of the density field, as this may help to improve the comprehension of the physical processes behind the formation and evolution of galaxies.

\begin{acknowledgements}
We acknowledge the financial contributions by grants ASI/INAF I/023/12/0 and PRIN MIUR 2010-2011 \textquotedblleft The dark Universe and the cosmic evolution of baryons: from current surveys to Euclid\textquotedblright. We thank the anonymous referee for the helpful comments. Reproduced with permission from Astronomy \& Astrophysics, \textcopyright ESO.
\end{acknowledgements}

\bibliographystyle{aa}
\bibliography{Malavasi2}

\end{document}